\documentclass[twocolumn]{aastex631}

\usepackage{amsmath,amssymb,amsfonts,amsthm}%
\usepackage{journal_abbreviations}

\usepackage{cleveref}

\usepackage[dvipsnames]{xcolor}
\begin{document}

\title{Ammonia-Hydrogen Phase Equilibria in Planetary Envelopes from First-Principles-Based Machine-Learning Molecular Dynamics}

\correspondingauthor{Akash Gupta; Jie Deng}
\email{akashgpt@princeton.edu; jie.deng@princeton.edu}

\author[0000-0002-2006-7769]{Akash Gupta}
\altaffiliation{51 Pegasi b Fellow}
\altaffiliation{Future Faculty in Physical Sciences Fellow}
\altaffiliation{Harry H. Hess Postdoctoral Fellow}
\affiliation{Department of Astrophysical Sciences, Princeton University, Princeton, NJ 08544, USA; {akashgpt@princeton.edu}}
\affiliation{Department of Geosciences, Princeton University, Princeton, NJ 08544, USA}

\author{Austin Guo}
\affiliation{Department of Astrophysical Sciences, Princeton University, Princeton, NJ 08544, USA}

\author[0000-0001-5441-2797]{Jie Deng}
\affiliation{Department of Geosciences, Princeton University, Princeton, NJ 08544, USA}

% \author{Adam Burrows}
% \affiliation{Department of Astrophysical Sciences, Princeton University, Princeton, NJ 08550, USA}



\begin{abstract}
Ammonia and hydrogen are major constituents of planets within and beyond the solar system, yet how they interact in planetary envelopes is poorly known. We perform large-scale two-phase coexistence simulations of the ammonia-hydrogen system from 500 to 4000 K and 0.1 to 100 GPa, driven by machine-learning interatomic potentials trained on density functional theory calculations. From these we map the compositions of coexisting hydrogen-rich and ammonia-rich phases, and the critical curve that separates conditions of complete miscibility from those of phase separation. The critical temperature increases with pressure, from 500 K at 0.1 GPa to 2000 K near 30 GPa, and we find no phase separation at 3000 and 4000 K up to 100 GPa. The slope of the critical curve decreases substantially between 0.3 and 30 GPa, and the curve nearly flattens above about 30 GPa and 2000 K, where molecules begin to give way to atomic species. The coexistence curves are asymmetric with an ammonia-rich critical composition. For the expected envelope compositions, the phase boundary lies close to, and intersects, published thermal profiles of Uranus and Neptune, implying ammonia may be raining out in these planets today. In contrast, in warm sub-Neptunes such as TOI-270 d, ammonia and hydrogen are likely miscible in all proportions. Together with previous studies, our results suggest that silicates, water and ammonia separate from hydrogen in sequence as planets cool, with implications for their compositional, thermal and structural evolution.
\end{abstract}

\keywords{Quantum-chemical calculations(2232) - Exoplanets(498) - Exoplanet atmospheres(487) - Mini Neptunes(1063) - Uranus(1751) - Neptune(1096) - Interdisciplinary astronomy(804) - Neural networks(1933) - Computational methods(1965) - Hot Neptunes(754)}
% atomistic simulations, exoplanets, Uranus, Neptune, planet atmospheres, planet interiors

%% and \citet commands to identify citations. The citations are

\section{Introduction}\label{sec:introduction}

Hydrogen is the main constituent of the envelopes of giant planets and of many sub-Neptunes, and ammonia, together with water and methane, is one of the planetary ``ices" thought to make up much of the interiors of Uranus and Neptune \citep[e.g.,][]{hubbard1980a}. Ammonia has been measured directly in the atmospheres of Jupiter and Saturn \citep[e.g.,][]{giles2017a,irwin2025a}. Beyond the solar system, ammonia has been proposed as a potential biosignature \citep{phillips2021a}, and its abundance in the atmospheres of warm sub-Neptunes such as K2-18 b and TOI-270 d has been used to constrain the presence of liquid-water oceans beneath them \citep{hu2021b,madhusudhan2023b}.

\begin{figure*}[ht!]
\centering
\includegraphics[width=\textwidth,trim=0 0 0 0,clip]{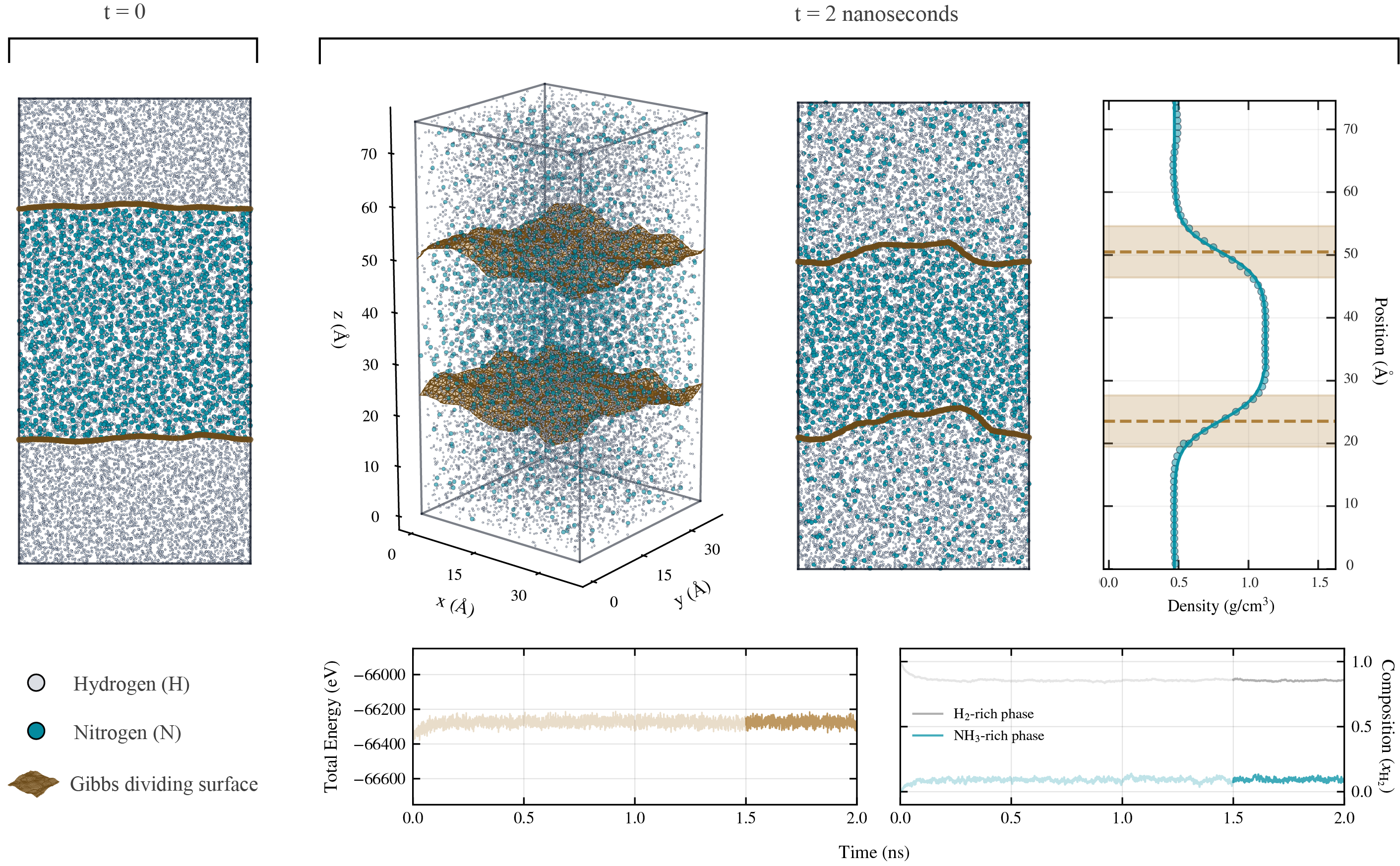}
\caption{A two-phase coexistence simulation at 1000 K and 7 GPa, containing 4096 H$_2$ and 2048 NH$_3$ molecules (16,384 atoms). Blue spheres are nitrogen atoms and gray spheres are hydrogen atoms. Top: the initial condition, in which slabs of pure ammonia and pure hydrogen are joined (left), followed by a perspective view and a side view of the system after 2 ns. The brown surfaces are the Gibbs dividing surfaces between the ammonia-rich and hydrogen-rich phases \citep{willard2010a}. The rightmost panel shows the mass density along the long axis of the cell after 2 ns; symbols are simulation results and the solid curve is the theoretically motivated hyperbolic-tangent profile expected for coexisting phases \citep{cahn1958a,widom1982a}. Dashed lines mark the determined interface positions and shaded bands reflect the interface widths. Bottom: total energy and composition of each phase as a function of time; the final 0.5 ns (darker) is the window over which the compositions are averaged.}
\label{fig:setup}
\end{figure*}

The phase behavior of pure ammonia has been studied extensively across planetary conditions, with experiments and first-principles calculations revealing molecular, ionic, superionic, and conducting states from sub-GPa pressures to hundreds of GPa and from hundreds to several thousand kelvin \citep[e.g.,][]{bethkenhagen2013a,ninet2012a,queyroux2019a,cavazzoni1999a,hernandez2023a}. In the envelopes of ice giants and sub-Neptunes, however, ammonia is mixed with hydrogen at conditions where hydrogen interacts strongly with other planetary materials. Recent theoretical, computational, and experimental studies have shown that hydrogen can react with, dissolve in, or separate from silicates and water at such conditions \citep[e.g.,][]{gupta2025a,gilmore2026a,horn2025a,bali2013a,rogers2025a,tejada2026a}. Whether ammonia and hydrogen mix or separate matters for these planets: separation releases gravitational energy as the denser phase sinks, and builds composition gradients that can inhibit convection \citep{stevenson1977a,leconte2012a}. Yet the ammonia-hydrogen system remains poorly constrained. Measurements exist only below the critical point of ammonia, at temperatures below about 400 K, where they find limited mutual solubility and extensive phase separation with increasing pressure \citep[e.g.,][]{wiebe1934a,wiebe1937a,krichevskii1939a,michels1950a,reamer1959a,bareli1961a,kaminishi1962a,heise1972a,moore1972a,ahrens1981a,chidester2011a}. Experiments at more extreme conditions of planetary envelopes and interiors are challenging.

Determining these phase equilibria at planetary conditions requires a method that remains accurate where molecules dissociate, ionize and react. Density functional theory (DFT) describes such materials without prior assumptions about their bonding, and DFT-based molecular dynamics (DFT-MD) has long been used to study planetary materials, including superionic water and ammonia \citep{cavazzoni1999a,bethkenhagen2013a}, the transport properties of hydrogen \citep{french2012a} and the phase separation of hydrogen and helium \citep{morales2009a,schottler2018a}. Simulating two coexisting phases, however, requires cells large enough to hold both phases and their interface, and runs long enough for their compositions to equilibrate, which for ammonia and hydrogen are beyond the reach of DFT-MD.

Machine-learning interatomic potentials (MLPs) trained on DFT energies, forces and stresses lift this limitation, retaining near-DFT accuracy at a cost that is orders of magnitude lower and grows linearly with the number of atoms, whereas the cost of DFT-MD grows as its cube \citep{behler2007a,zhang2018a}. They have been used, for example, to compute the phase diagram of water \citep{zhang2021a} and the melting curve of MgSiO$_3$ \citep{deng2023b}. A potential that is reliable over a wide range of conditions must, however, be trained on configurations of every relevant phase and of the interfaces between them, which simulations at single state points rarely visit. We therefore build our MLP using active learning, combining enhanced sampling, descriptor-based selection of configurations and DFT recalculation (Appendix~\ref{sec:methods}).

With this potential, we simulate coexisting ammonia and hydrogen mixtures in cells of 16,000-70,000 atoms for up to several nanoseconds, at 500-4000 K and 0.1-100 GPa. These simulations map where ammonia and hydrogen form a single homogeneous phase and where they separate into two phases of distinct composition. From them we determine the coexistence (binodal) surface and the critical curve, which nearly flattens above about 30 GPa and 2000 K, where molecules begin to give way to atomic species. Comparison with models of Uranus and Neptune shows that ammonia may separate from hydrogen in their envelopes today, at depths that depend on the interior model. In warm sub-Neptunes such as TOI-270 d, by contrast, ammonia and hydrogen are likely miscible in all proportions.

\begin{figure*}
\centering
\includegraphics[width=\textwidth,trim=5 0 0 80,clip]{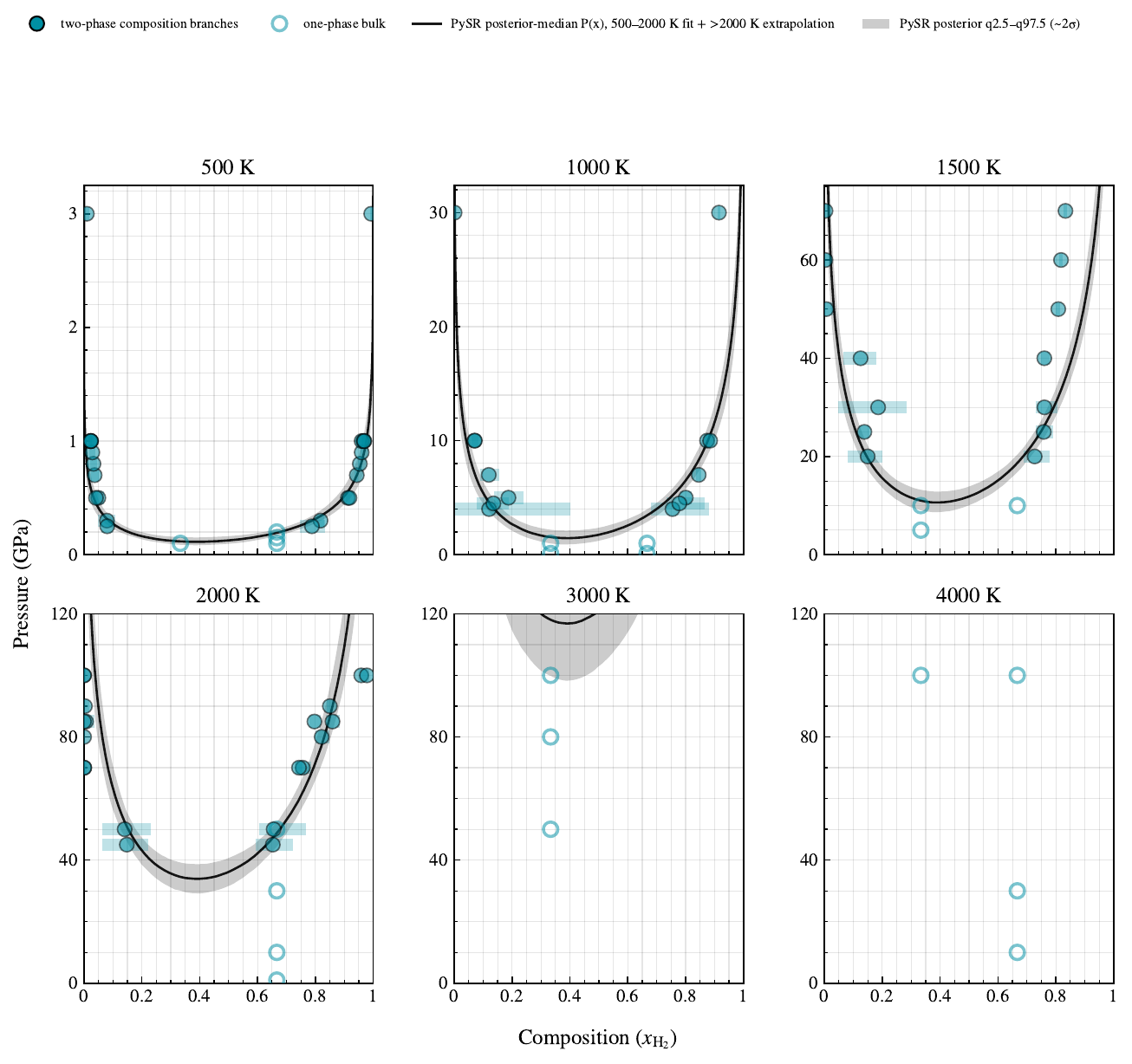}
\caption{Compositions of coexisting phases as a function of pressure at six temperatures. Filled circles are the measured compositions of each phase in simulations that result in two phases, with bands indicating their 2$\sigma$ variation in time, and open circles show the compositions of simulations that remain a single homogeneous fluid. The solid curves are the posterior-median coexistence curves from our thermodynamic model, with shaded 2$\sigma$ envelopes (see \textit{Methods}). The simulations have bulk compositions of $x_{\text{H}_2}=$ 0.67 and 0.33, and contain between 16,384 and 70,304 atoms. No simulation at 3000 or 4000 K separates into two phases at any pressure that we sampled and thus the coexistence curve shown at 3000 K is only partially constrained.}
\label{fig:coexistence}
\end{figure*}

\section{Results}\label{sec:results}

\subsection{Phase coexistence and separation}

We use large-scale phase coexistence MD simulations to determine phase equilibria in the NH$_3$-H$_2$ system. These simulations are driven by machine-learning interatomic potentials (MLPs) trained on high-fidelity data from DFT calculations. Under this framework, each simulation starts from two equilibrated slabs of fluid, one of pure NH$_3$ and one of pure H$_2$, placed in contact across a planar interface \citep[left panel, Figure \ref{fig:setup};][]{gupta2025a,insixiengmay2025a,gilmore2026a}. This setup is then allowed to evolve for several nanoseconds until energy and compositional equilibrium is reached (bottom panels, Figure \ref{fig:setup}). We then perform statistical analysis of the last 0.5 nanoseconds of the simulation.

Our simulations cover temperatures of 500-4000 K and pressures of 0.1-100 GPa (1 to 10$^3$ kbar). Across these P-T conditions, we find ammonia and hydrogen either exist as one homogeneous fluid or phase-separate into two compositionally distinct phases existing in dynamic equilibrium: one richer in hydrogen and the other in ammonia. Mapping the compositions $x$ of the homogeneous or coexisting phases across temperatures (T) and pressures (P) allows us to estimate the critical curve which delineates the P-T-$x$ stability regime of single homogeneous fluid of ammonia and hydrogen.

We express the composition of each phase as its mole fraction of hydrogen
\begin{equation}
    x_{\text{H}_2}=\frac{n_{\text{H}_2}}{n_{\text{H}_2}+n_{\text{NH}_3}}=\frac{n_\text{H}-3n_\text{N}}{n_\text{H}-n_\text{N}},
\end{equation}
in which $n_i$ counts the atoms or molecules of species $i$. The two coexisting phases are identified with a three-dimensional Gibbs dividing surface constructed from a coarse-grained density field over the analysis window.

Figure \ref{fig:coexistence} shows the compositions of the coexisting phases as a function of pressure at six temperatures. A coexistence or binodal curve defines the boundary between the single-phase and two-phase regions in the composition-pressure space at some temperature. A continuum of these across temperatures is defined as a coexistence or binodal surface. This figure shows that at the lowest pressure end of 0.1 GPa, hydrogen and ammonia are completely miscible across the explored T-P regime. As the pressure increases, the fluid mixture becomes unstable and separates into two phases whose compositions diverge. At a given temperature, the two coexistence branches merge at the critical pressure P$_c$ and critical composition $x_c$. The exceptions are the two highest-temperature isotherms we explored: 3000 and 4000 K. At these temperatures, we do not find any demixing even at pressures of $\sim$ 100 GPa which is on the higher end of possible pressures at the atmosphere-interior boundary \citep{nettelmann2013a}. We describe this phase diagram with an asymmetric (subregular) solution model whose mixing parameters depend on temperature (Methods).

We find the coexistence curves between 500 to 2000 K are asymmetric, with an ammonia-rich critical composition of $x_c$ $\sim$ 0.4. In other words, at a certain P-T condition, an ammonia-rich phase has less hydrogen dissolved in it in comparison to the amount of ammonia that could dissolve into the hydrogen-rich fluid.

\begin{figure*}
\centering
\includegraphics[width=\textwidth,trim=0 0 0 0,clip]{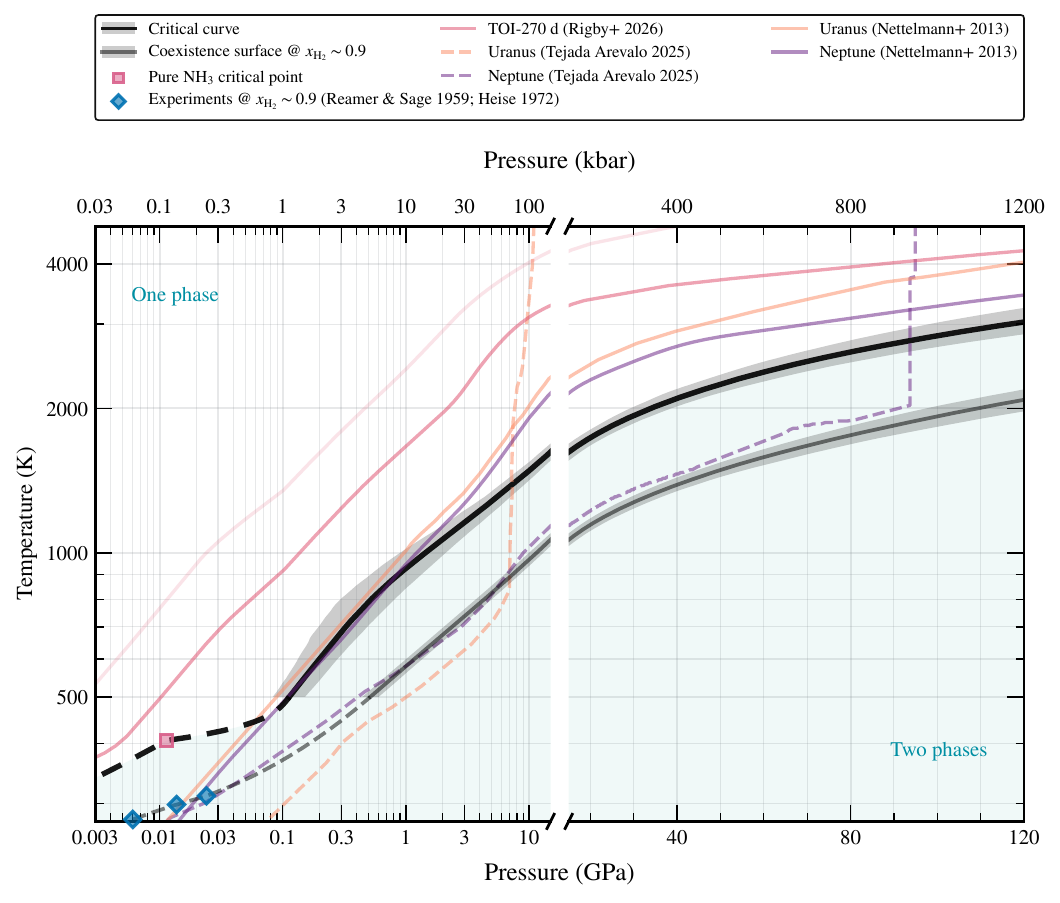}
\caption{Phase diagram of the ammonia (NH$_3$)-hydrogen (H$_2$) system in pressure and temperature space. The bold black line with its gray 2$\sigma$ band is the critical curve from our thermodynamic model delineating the regimes of phase stability (turquoise shading). Above 2000 K the critical curve is only partially constrained because none of the simulations at 3000-4000 K phase-separate up to 100 GPa. The bold dashed line below 500 K is physically motivated by previous work on hydrogen-water fluids \citep{seward1981a,gupta2025a} and joins our results to the critical point of pure ammonia (pink square) and its condensation curve \citep{gao2023a}. The gray solid line, with its 2$\sigma$ band, denotes the isopleth of the coexistence surface at $x_{\text{H}_2}\sim0.9$, as expected in the deep envelopes of Uranus and Neptune \citep{nettelmann2013a,scheibe2019a}. It is extended to experimental results from \citet{reamer1959a} and \citet{heise1972a}. Also shown are model P-T profiles for Uranus (orange) and Neptune (purple) from \citet{nettelmann2013a} (solid lines) and  \citet{tejada2025a} (dashed lines). The red lines are P-T profiles of TOI-270 d from \citet{rigby2026a} for internal temperatures of 25 K (dark red) and 50 K (light red). The pressure axis is logarithmic at low pressure and linear at high pressure. Comparison of the critical curve and this isopleth with the model profiles of Uranus, Neptune, and TOI-270 d suggests that while ammonia and hydrogen are likely fully miscible in the envelopes of warm sub-Neptunes, Uranus and Neptune may be undergoing ammonia rain today, at depths that depend on assumptions about the initial conditions and evolution of these planets.}
\label{fig:critical}
\end{figure*}

In Figure \ref{fig:critical}, we show how the critical curve (black solid line) of the NH$_3$-H$_2$ system changes across the temperature-pressure space. The critical curve separates the conditions under which ammonia and hydrogen mix in all proportions as a single homogeneous fluid from those under which they can separate into two phases, one richer in hydrogen and the other in ammonia, traced by the two branches of the coexistence curves (Figure \ref{fig:coexistence}). At each temperature it gives the critical pressure P$_c$ at which these two branches merge, or equivalently, at each pressure, the critical temperature T$_c$ above which the two fluids mix in all proportions. We find that the critical pressure rises monotonically with temperature. From a critical pressure of $\sim$0.1 GPa at 500 K, it increases to $\sim$30 GPa at 2000 K. The top panel in Figure \ref{fig:speciation} demonstrates that the critical curve slope $d \text{T}_c/d \text{P}_c$ decreases continuously with increasing pressure: from $\sim$700 K/GPa at 0.3 GPa to $\sim$20 K/GPa at 30 GPa, where T$_c$ $\sim$ 2000 K. Beyond $\sim$30 GPa, the critical curve flattens to a gradient of $\sim$10 K/GPa at 100 GPa. The critical pressure consequently rapidly rises beyond 100 GPa above 3000 K.

In Figure \ref{fig:critical}, we also show as a gray line the isopleth of the coexistence surface at $x_{\text{H}_2}\sim0.9$, that is, the boundary of two-phase stability for a mixture of this fixed composition. This is motivated by constraints on the bulk metallicity of the envelopes of Uranus and Neptune, $Z\sim$0.8-0.9 \citep{nettelmann2013a,scheibe2019a}. Assuming an O:C:N ratio of 7:4:1 \citep{asplund2009a}, this $Z$ implies $x_{\text{H}_2}\sim0.9$ for the NH$_3$-H$_2$ binary mixture.

Finally, Figure \ref{fig:critical} compares the critical curve with model P-T profiles of planets with hydrogen-rich envelopes. The profiles of \citet{nettelmann2013a} cross the extended critical curve at $\sim$0.07 GPa and then lie within its 2$\sigma$ band up to $\sim$1-2 GPa; they cross the $x_{\text{H}_2}\sim0.9$ isopleth at $\sim$0.02 GPa (200 bar). The colder profiles of \citet{tejada2025a} lie close to or below this isopleth down to $\sim$7 GPa. The profiles of TOI-270 d from \citet{rigby2026a} lie well within the one-phase stability region. It should be noted though that this is not a self-consistent comparison as these planet modeling studies did not account for phase separation within the envelopes and its energetic and structural implications for the planet's evolution.

\subsection{Molecular-to-atomic transition}
To understand these changes in the critical curve, we analyze the structure and bonding of homogeneous NH$_3$-H$_2$ mixtures, simulated along the critical curve up to 3000 K and above it at 4000 K at the highest pressures explored. 

\begin{figure*}
\centering
\includegraphics[width=\textwidth,trim=5 0 5 20,clip]{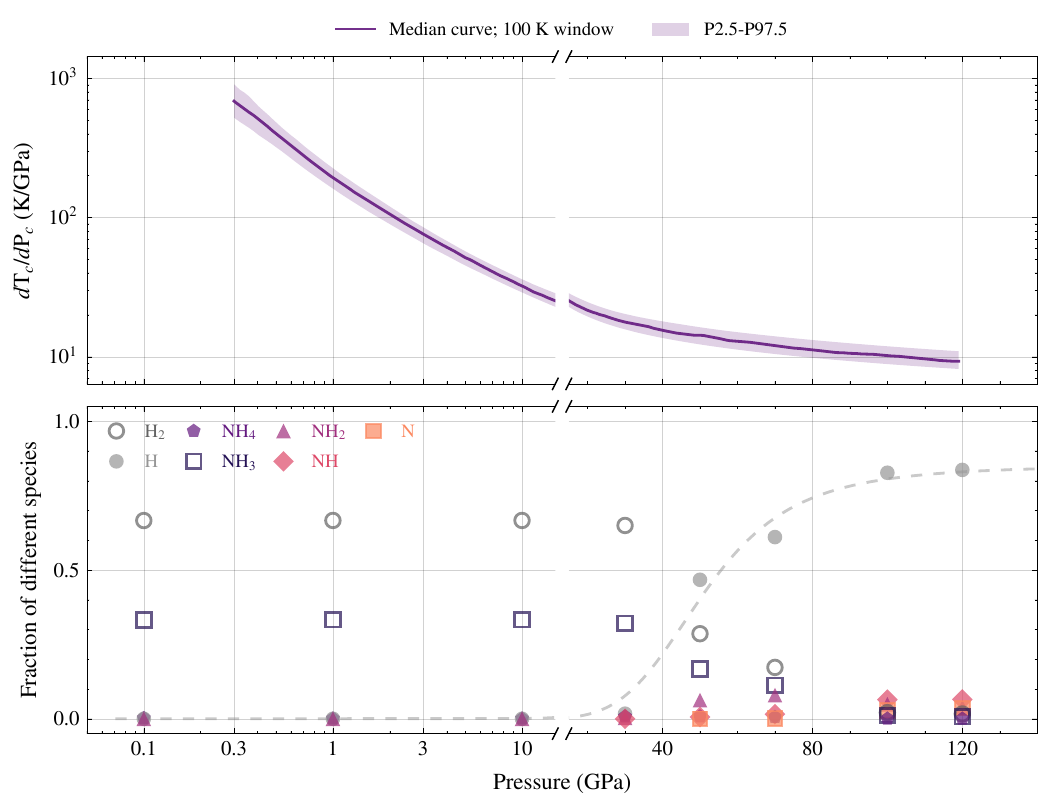}
\caption{Variation with pressure of the slope of the critical curve (top) and of the speciation of the homogeneous fluid (bottom). In the top panel, the line shows the slope $d\mathrm{T}_c/d\mathrm{P}_c$ of the critical curve, and the shaded band its 2$\sigma$ uncertainty. The bottom panel gives the fraction of each species in a homogeneous fluid with $x_{\text{H}_2}=0.67$ along the critical curve up to 3000 K and then at 4000 K at the highest pressures. The gray dashed curve is a Fermi-Dirac-like function tracing the change in atomic H fractions, showcasing the molecular-to-atomic transition of the fluid at about 30-100 GPa and 2000-4000 K. Species are identified from atomic neighbors that persist for 100 fs. The pressure axis is logarithmic below 15 GPa and linear above.}
\label{fig:speciation}
\end{figure*}

We find that the flattening of the critical curve follows the molecular-to-atomic transition of the fluid (bottom panel, Figure \ref{fig:speciation}). Below 2000 K and 10 GPa, the fluid is primarily molecular H$_2$ and NH$_3$, with trace amounts of NH$_4$. Between 2000 K, 10 GPa and 3000 K, 100 GPa, we see a gradual increase in the fraction of atomic H and single-, double- and triple-dissociated ammonia species, namely, NH$_2$, NH, N. Degree of dissociation for NH$_3$ also rises with increasing temperature and pressure along the critical curve. At $\lesssim$ 2000 K nearly all nitrogen atoms are in NH$_3$ molecules. At 3000 K the fraction of nitrogen atoms in NH$_3$ falls to about 70 and 50 percent at 50 and 70 GPa, and at 4000 K and 100-120 GPa it falls to about 5 percent. These are fractions of nitrogen atoms for a 100 fs persistence window (Methods), whereas Figure \ref{fig:speciation} shows the fractions of species. NH$_2$ becomes prominent at 3000 K, while NH and N account for larger nitrogen fractions at 4000 K.

The observed molecular-to-atomic transition of H occurs at conditions similar to those of the molecular-to-atomic transition in pure hydrogen \citep[e.g.,][]{mcwilliams2016a} and in first-principles work on other H-bearing systems such as water-hydrogen mixtures \citep{gupta2025a}. Similarly, the dissociation of nitrogen species is consistent with past laboratory experiments and theoretical calculations \citep{bethkenhagen2013a,ravasio2021a}.

\section{Discussion}\label{sec:discussion}

Our results delineate the conditions under which ammonia and hydrogen mix in all proportions and when they rather separate into two phases. The critical curve and the coexistence surface that bound these conditions run close to, and in places cross, the P-T profiles inferred for the envelopes of Uranus and Neptune, but lie well below those of warm sub-Neptunes (Figure \ref{fig:critical}). Whether and at what depth ammonia separates from hydrogen in a planet therefore depends on its temperature profile and bulk composition. This has implications for the compositional and thermal evolution of such planets, and for connecting observations of their atmospheres to their deep envelopes and interiors.

\subsection{Phase separation in Uranus and Neptune}

In the early lives of Uranus and Neptune, their envelopes were hotter than they are today. Despite their cold equilibrium temperatures of $\sim$50 K \citep{pearl1990a,pearl1991a}, much of their P-T profiles would then have lain above the NH$_3$-H$_2$ critical curve, so that ammonia and hydrogen mixed in all proportions except in the outermost atmosphere, where ammonia condenses into clouds. As the planets cooled, their temperatures fell at every pressure, so the levels at which ammonia and water condense, and with them their cloud decks, moved deeper, as is general for condensate clouds \citep{pierrehumbert2010a,burrows2006a}.

Because ammonia is far less volatile than hydrogen, the two-phase region of NH$_3$-H$_2$ mixtures is bounded by two curves that meet at the critical point of ammonia (405.4 K, 11.3 MPa). Above it, the bound is the critical curve, along which the two coexisting fluids become identical. Below it, the bound is the condensation curve of pure ammonia: all measured vapor-liquid equilibria of NH$_3$-H$_2$, at 278–394 K and up to 0.04 GPa for the gas phase and 0.1 GPa for the liquid, lie on its two-phase side \citep{wiebe1934a,reamer1959a,heise1972a}. The same holds for H$_2$O-H$_2$, whose critical curve begins at the critical point of water \citep{seward1981a}. Crossing this bound on cooling is necessary for phase separation but not sufficient: a mixture separates only below the isopleth of its own composition on the coexistence surface (Figure \ref{fig:coexistence}). This coexistence surface describes vapor-liquid equilibrium below the critical point of ammonia and fluid-fluid equilibrium above it. In Figure \ref{fig:critical}, the black line is its outer bound and the gray line its cross-section at $x_{\text{H}_2}\sim0.9$. Below 500 K, both rest on previous measurements and the condensation curve of ammonia rather than on our simulations.

In present-day Uranus and Neptune, comparison with the model profiles of \citet{nettelmann2013a} suggests that ammonia and hydrogen may phase separate at pressures below $\sim$0.02 GPa (200 bar), and are miscible in all proportions beyond $\sim$1-2 GPa, with the profiles within the uncertainty of the critical curve in between. The colder, non-adiabatic profiles of \citet{tejada2025a} instead lie close to or below the $x_{\text{H}_2}\sim0.9$ isopleth down to $\sim$7 GPa. Ammonia could thus separate from hydrogen only in the outermost $\sim$200 bar of the hotter adiabatic models of \citet{nettelmann2013a}, where it would condense into a cloud deck, but down to $\sim$7 GPa in the non-adiabatic models \citep{tejada2025a}. Whether and how deep ammonia separates today therefore depends on the assumptions of the envelope-interior models. As the planets cool and this level deepens, separation would gradually deplete ammonia in the outer envelope. Direct observation of this depletion is complicated, however, because much of the ammonia at a few to tens of bar is bound with H$_2$S as NH$_4$SH in clouds, as noted in previous studies \citep{atreya2020a,molter2021a,tollefson2021a}.

Where the mixture does phase-separate, the ammonia-rich phase ($x_{\text{NH}_3}\gtrsim0.9$), being denser (Figure \ref{fig:setup}), rains out and sinks. This releases gravitational energy and latent heat, in addition to that from the separation of water \citep{bailey2021a,gupta2025a}, much as helium rain is thought to contribute to the luminosity of the gas giants \citep{stevenson1977a,stevenson1977b}. The composition gradient that it builds could also inhibit convection \citep{leconte2012a}. Phase separation of ammonia may therefore matter for the thermal and radius evolution of Uranus and Neptune, and of cold exoplanets of similar mass, discussed below.

\subsection{Consequences of the molecular-to-atomic transition}

We find the molecular-to-atomic transition to occur at about 30-100 GPa and 2000-4000 K. This is in a similar range as in hydrogen \citep[e.g.,][]{mcwilliams2016a} and in H$_2$O-H$_2$ fluids \citep{gupta2025a}, and lies close to the conditions along model profiles of Uranus and Neptune (Figure \ref{fig:critical}). In our simulations, the substantially dissociated states are all single-phase. A fluid of atomic hydrogen and nitrogen-bearing fragments is a different material from one of H$_2$ and NH$_3$ molecules. Its thermal and transport properties are therefore expected to differ, as they do in hydrogen \citep{french2012a} and water \citep{french2019a,grasselli2020a}. In particular, dissociated hydrogen, water and ammonia conduct electricity and heat better than their molecular forms \citep{cavazzoni1999a,ravasio2021a,french2012a,grasselli2020a}. Where dissociation is under way, part of the energy that would otherwise raise the temperature goes into breaking bonds, which lowers the adiabatic temperature gradient \citep{saumon1995a,french2012a}. The gradient rises again as the fluid recombines on cooling. These properties govern whether heat is carried by convection or by conduction \citep{scheibe2021a}, and where the magnetic field is generated \citep{stanley2004a}.

\subsection{Exoplanets}

Most known exoplanets are much hotter than the ice giants \citep{bean2021a}. Like those of TOI-270 d \citep[Figure \ref{fig:critical};][]{rigby2026a}, their P-T profiles lie on the one-phase side of the critical curve. In such planets ammonia, like water \citep{gupta2025a}, could be fully miscible with hydrogen, so that no phase boundary separates the deep nitrogen inventory from the observable atmosphere. A mixed envelope implies a higher mean molecular weight than one of pure hydrogen, consistent with the metal-rich atmosphere inferred for TOI-270 d \citep{benneke2024a}. 

Our results suggest that NH$_3$–H$_2$ phase separation is thus unlikely to explain the low inferred atmospheric abundance or non-detection of ammonia in such planets. Other proposed explanations include low-pressure thermochemical conversion of NH$_3$ to N$_2$ followed by vertical quenching \citep{ohno2023a}, photochemical destruction of NH$_3$ \citep{hu2021a,ohno2023b}, and nitrogen sequestration into liquid-water oceans \citep{hu2021b} or magma oceans \citep{shorttle2024a}. The relative importance of these mechanisms remains debated \citep{madhusudhan2023a,wogan2024a}.

Cold exoplanets of Neptune mass and below may instead resemble Uranus and Neptune. Microlensing surveys show that such planets are common, for example OGLE-2007-BLG-368Lb \citep{sumi2010a}, MOA-2009-BLG-266Lb \citep{muraki2011a} and OGLE-2012-BLG-0950Lb \citep{koshimoto2017a}, and the Nancy Grace Roman Space Telescope is expected to find $\mathcal{O}(1000)$ planets with this technique \citep{penny2019a}. If their deep envelopes are as cold as those of the ice giants, ammonia and water may separate from their hydrogen-rich atmospheres, which will have implications for their thermal evolution.

\subsection{Ammonia versus water}

Water is the other major ``ice" in Uranus and Neptune and may be abundant in many sub-Neptunes \citep{bean2021a}, and its miscibility with hydrogen has been studied extensively \citep{seward1981a,bali2013a,soubiran2015a,bailey2021a,canoamoros2024a,howard2025a,gupta2025a,bergermann2024a}. The critical curves of the H$_2$O-H$_2$ \citep{gupta2025a} and NH$_3$-H$_2$ systems are similar up to a few GPa. At higher pressures, the NH$_3$-H$_2$ critical curve is the colder of the two, by several hundred Kelvin at 10-30 GPa. 

The bulk abundance of water may, however, be about seven times that of ammonia \citep{asplund2009a}. For a Neptune-like envelope, the bulk composition of the H$_2$O-H$_2$ binary, $x_{\text{H}_2}\sim 0.6$-0.7, therefore lies close to its critical composition of roughly 0.6-0.7 \citep{gupta2025a}, whereas that of the NH$_3$-H$_2$ binary, $x_{\text{H}_2}\sim0.9$, lies far from its critical composition of $\sim0.4$. Water thus separates close to its critical curve, whereas ammonia requires temperatures well below its own, so that at a given temperature water separates at lower pressure than ammonia. In a planet that contains both, water and ammonia therefore need not separate from hydrogen at the same depth or at the same time. The ice giants illustrate this: the hotter interior models of Uranus and Neptune \citep{nettelmann2013a} lie close to both critical curves at 10-30 GPa, where water may be separating, or starting to, whereas in the colder model of \citet{tejada2025a} water would have largely separated.

Where both are present, however, ammonia could separate earlier than our phase diagram suggests if, being polar, it partitions into the water-rich fluid as water separates \citep{bailey2021a}. Other calculations suggest that at high pressures, water instead separates from ammonia and methane \citep{militzer2024a}. This ammonia-water partitioning with temperature-pressure conditions is thus uncertain and requires study of the ternary system. Resolving it matters for interpreting the tropospheres of Uranus and Neptune, which are poor in nitrogen relative to sulfur \citep{molter2021a,tollefson2021a}, an observation that has been attributed to nitrogen-poor accretion or to uptake of ammonia by a deep water-rich or ionic layer \citep{tollefson2021a,atreya2020a}. It would likewise affect how much nitrogen the envelopes of cold exoplanets retain.

\subsection{Limitations and outlook}

The inferences here are subject to certain caveats. In particular, our computational experiments treat the binary system of ammonia and hydrogen, whereas planetary envelopes also contain helium, water, methane and other materials. The critical curve above 2000 K is only partially constrained: we find no two-phase state at 3000 and 4000 K up to 100 GPa, so that comparisons with planetary profiles at pressures above about 30 GPa are dependent on our thermodynamic model. 

Published envelope-interior models of the ice giants also differ by several hundred kelvin at the same pressure (Figure \ref{fig:critical}), so that our conclusions for present-day Uranus and Neptune depend on the assumptions of the model adopted. Our comparisons also assume local thermodynamic equilibrium, whereas atmospheric dynamics may shape cloud levels and the distribution of ammonia \citep{li2017a,guillot2020a,molter2021a,tollefson2021a}. Future work should couple these phase relations to models of thermal evolution and vertical transport, including the energetics of phase separation and the molecular-to-atomic transition. Despite these uncertainties, our results demonstrate that the miscibility of ammonia with hydrogen, like that of water and silicates \citep{gupta2025a,gilmore2026a}, changes within the range of conditions found in the envelopes of ice giants and sub-Neptune-to-Neptune class exoplanets: phase behavior thus needs to be taken into account in understanding the compositional, thermal and structural evolution of these planets.

\section{Conclusions}\label{sec:conclusions}

In this work, we determined the phase diagram of NH$_3$-H$_2$ mixtures at 500-4000 K and 0.1-100 GPa. The critical curve, which divides conditions under which the mixture is a single homogeneous fluid from those under which it separates into two phases, rises from 500 K at 0.1 GPa to 2000 K near 30 GPa. It flattens with increasing pressure, a change that broadly overlaps with the molecular-to-atomic transition of the fluid, which is expected to alter its thermal and transport properties. The coexistence curves are asymmetric, with an ammonia-rich critical composition (x$_c\sim$0.4). Together with previous measurements at lower pressures, our results are consistent with a single, continuous coexistence surface that spans vapor-liquid equilibrium below the critical point of ammonia, where ammonia condenses into clouds, and fluid-fluid equilibrium above it, where an ammonia-rich fluid separates from hydrogen at depth.

In warm sub-Neptunes such as TOI-270 d, ammonia and hydrogen are expected to be miscible in all proportions throughout the envelope, so that a low inferred atmospheric abundance or non-detection of ammonia in these planets more likely has other causes. Uranus and Neptune, in contrast, may be experiencing ammonia rain today, although whether and how deep ammonia separates depends on the envelope-interior models and their inferred compositions. For instance, separation could be confined to the outermost $\sim$200 bar in hotter adiabatic models but extend to GPa depths in colder, non-adiabatic ones. Water, whose bulk composition lies closer to its critical composition, likely separates at lower pressures than ammonia, so the two need not separate at the same depth or time, although partitioning of ammonia into a water-rich fluid could couple them.

Our results provide the phase relations needed to include ammonia in models of the compositional, thermal and structural evolution of hydrogen-rich planets. Where ammonia separates, the denser ammonia-rich phase sinks, releasing energy that, together with the separation of water, may influence the thermal and radius evolution of the ice giants and of cold Neptune-mass exoplanets. The compositions of the coexisting phases also set how nitrogen is distributed between the envelope and the interior as a planet cools, so that a planet's present nitrogen distribution may reflect its thermal evolution as well as the composition it acquired during formation. Together with recent studies of water and silicates \citep{gupta2025a,gilmore2026a}, our results suggest that separation of different species from hydrogen is likely common in Neptune-like planets and proceeds in sequence, with less volatile components such as silicates separating first and more volatile ones such as water and ammonia following it, as a planet cools over billions of years or across planets of decreasing equilibrium temperature.

% \begin{acknowledgments}
\section*{Acknowledgments}
A.G. is grateful to Adam Burrows for helpful discussions, insights, and access to computational resources at Princeton University. A.G. acknowledges support from the Heising–Simons Foundation through the 51 Pegasi b Fellowship, and from Princeton University through the Harry H. Hess Fellowship and the Future Faculty in Physical Sciences Fellowship. A.G. also acknowledges support from the National Aeronautics and Space Administration (NASA) under grant HST-GO-17804.011-A and the Center for Matter at Atomic Pressures (CMAP), a National Science Foundation (NSF) Physics Frontier Center under Award PHY-2020249. J.D. acknowledges support from the NSF under Award EAR-2242946.

The calculations presented in this article were performed on computational resources of the Argonne Leadership Computing Facility, which is a U.S. Department of Energy Office of Science User Facility operated under contract DE-AC02-06CH11357, and those managed and supported by Princeton Research Computing, a consortium of groups including the Princeton Institute for Computational Science and Engineering (PICSciE) and the Office of Information Technology's High Performance Computing Center and Visualization Laboratory at Princeton University.
% \end{acknowledgments}

\software{VASP \citep{kresse1996a,kresse1999a}, DeePMD-kit \citep{wang2018a,zhang2018a}, LAMMPS \citep{thompson2022a}, PLUMED \citep{tribello2014a,Bonomi2009a,Bonomi2019a}, ALCHEMY, pytim \citep{sega2018a}, PySR \citep{cranmer2023a}, Nautilus \citep{lange2023a}, NumPy \citep{harris2020a} and Matplotlib \citep{hunter2007a}.}

\newpage

\appendix
\section{Methods}\label{sec:methods}

\subsection{Computation}

To study the NH$_3$-H$_2$ system, we perform large-scale molecular dynamics (MD) simulations driven by a machine-learning interatomic potential (MLP) that is trained on high-fidelity electronic structure calculations. This methodology is divided into two stages. Stage 1 consists of developing an MLP based on atomic interactions across the relevant temperature, pressure, compositional, and configurational space. Stage 2 then involves large-scale machine learning molecular dynamics (MLMD) simulations at near-DFT-accuracy, driven by the curated MLP.

For stage 1, we leverage \textsc{ALCHEMY} (Ab initio-deep Learning Coupled Hybrid Engine for planetary Materials Exploration and discoverY), an automated active-learning pipeline that involves four phases: training, exploration, selection and labeling, to develop a balanced, DFT-accuracy-level MLP \citep{deng2023b,ZhangYuzhi2020a}. For the first phase within stage 1, we adopt the smoothed version of Deep Potential \citep{zhang2018a,zhang2018b} as implemented in DeePMD-kit \citep{wang2018a,Zeng2023a,Zeng2025a}: a deep learning package for training interatomic interactions based on ab initio data. This approach does not prescribe an explicit analytic functional form for the potential energy surface or force field, and rather uses a deep neural network to model it. The total energy is written as the sum of atomic contributions, each of which depends on the local environment of the atom within a cutoff radius of 6.0 \AA{}, smoothed from 5.8 \AA{}. The descriptor of the local environment is parameterized by an embedding network with three hidden layers of 25, 50 and 100 nodes, and the energy is obtained from a fitting network with five hidden layers of 300 nodes each. The electronic temperature is included as an additional frame parameter, which allows the potential to distinguish the energies and forces of identical configurations at different temperatures, so that a single model describes the entire temperature range of this study. The MLP is trained on energies, forces, and virials with an initial learning rate of 0.001, which decays exponentially every 10,000 steps.

In each iteration, during the second phase, an MLP trained on all DFT data accumulated until that point drives MD simulations using the Large-scale Atomic/Molecular Massively Parallel Simulator \citep[LAMMPS;][]{thompson2022a} package that explore pressure-temperature  (P-T) regimes across compositions from pure ammonia to $x_{\text{H}_2}$=0.99, in cells of 128-288 atoms. The exploration either ramps temperature and pressure in the isothermal-isobaric ensemble (NPT), or uses variationally enhanced sampling \citep{valsson2014a} with PLUMED \citep{tribello2014a,Bonomi2009a,Bonomi2019a}. In the latter, a bias on the potential energy, the volume, and a ratio of N-N to N-H coordination numbers that distinguishes mixed from demixed states is optimized to sample the multithermal-multibaric distribution in these three variables \citep{piaggi2019a,piaggi2019b}, so that a simulation visits mixed and demixed configurations over the whole temperature and pressure range of a regime. Each such simulation is run for 0.2-0.5 ns with a time-step of 0.5 fs. Across all iterations, this amounts to a total of 1.2 $\mu$s of simulation data.

In the third phase, we select mutually dissimilar configurations from each trajectory by farthest-point sampling in the space of Smooth Overlap of Atomic Positions (SOAP) descriptors \citep[SOAP; ][]{bartok2013a,cheng2020a}. The selected frames are labeled with DFT, and those for which the MLP error in energy, force or virial lies between prescribed lower and upper thresholds are added to the training set. The lower threshold is set by the required MLP accuracy, whereas the upper threshold excludes configurations that are likely to be unphysical. A composition is retired from a regime once fewer than 5\% of its new labels exceed the lower threshold. Driver scripts automate the loop by submitting and monitoring the training, MD and DFT jobs on the computing clusters, adapting the lower threshold, and starting the next iteration without user intervention. The regimes explored span 500-6500 K and 0.05-300 GPa. The final MLP, used for all results reported here, is the outcome of 44 iterations and is trained for 1.5 million optimization steps with Adam optimizer \citep{kingma2015a} on 68,994 configurations containing 16,128,230 atoms.

The labeling phase, i.e., the electronic structure calculations based on DFT that determine the reference energies, forces and virials of atomic configurations, is performed with the Vienna \textit{ab initio} Simulation Package \citep[VASP;][]{kresse1993a,kresse1996a,kresse1996b,kresse1999a} based on the projector-augmented wave (PAW) method \citep{blochl1994a}. H and N have valence configurations of 1s$^1$ and 2s$^2$2p$^3$, with PAW core radii of 1.10 and 1.50 bohr (0.582 and 0.794 \AA{}), respectively. The regularized-restored Strongly Constrained and Appropriately Normed (r$^2$SCAN) functional \citep{furness2020a} is used for the exchange-correlation energy, and the non-spherical contributions to the gradient corrections inside the PAW spheres are included. This nonempirical meta-generalized gradient approximation (meta-GGA) is built from exact physical constraints and retains much of the transferable accuracy of SCAN \citep{sun2015a} while improving its numerical stability and efficiency. It therefore offers a practical balance of accuracy and computational cost for the large number of reference calculations required here. Benchmarks demonstrate good performance for a range of properties in both solids and fluids, including fluid conditions from ambient to supercritical \citep{kingsbury2022a,chattopadhyay2026a}. Consistent with this transferability, the parent SCAN functional yields a water phase diagram spanning molecular and ionic fluids and numerous ice polymorphs in satisfactory overall agreement with experiment \citep{sun2015a,zhang2021a}. The kinetic-energy cutoff of the plane waves is 800 eV, and the Brillouin zone is sampled at the Gamma point. For five configurations with $x_{\text{H}_2}=0.5$ at 1000 K and 0-20 GPa, these settings yield energies, forces and pressures that are within 0.05 meV atom$^{-1}$, 19 meV \AA$^{-1}$ and 0.02 GPa, respectively, of those obtained with a $2\times2\times2$ k-point mesh and a cutoff of 1200 eV. The self-consistent field iteration stops when the changes in total energy and in band-structure energy are less than $10^{-5}$ eV. The electronic states are occupied according to the Fermi-Dirac distribution at an electronic temperature equal to the ionic temperature, i.e., we assume thermal equilibrium between ions and electrons via the Mermin functional \citep{mermin1965a}, and the MLP is trained on the resulting free energies inclusive of electronic entropy.

Over the training set, the root-mean-square errors of the final MLP are 12 meV atom$^{-1}$ in energy, 0.48 eV \AA$^{-1}$ in force and 2.38 GPa in stress (28 meV atom$^{-1}$ in virial). The errors in force and stress grow with temperature and pressure (Figure \ref{fig:training-errors}): the mean error in force rises from 0.23 eV \AA$^{-1}$ at 500 K (0.05-160 GPa) to 0.58 eV \AA$^{-1}$ at 4000 K (0.05-300 GPa), and that in stress from 0.21 GPa to 1.64 GPa, which is expected since higher temperatures and pressures result in larger fluctuations of force and stress. The mean error in energy varies little with temperature, between 11 and 15 meV atom$^{-1}$. We test the final MLP against DFT on 258 configurations at 500, 1000, 2000 and 4000 K that are distinct from the training set (Figure \ref{fig:parity}, top panels). This test set contains 61,562 atoms in total, and therefore 258 energies, 184,686 force components and 1,548 stress components. The root-mean-square errors are 19 meV atom$^{-1}$ in energy, 0.39 eV \AA$^{-1}$ in force and 0.96 GPa in stress. Furthermore, we find radial distribution functions from MLMD agree well with those from DFT-MD at 1000 K and 30 GPa (Figure \ref{fig:parity}, bottom left). 

We then leverage this final, curated MLP to perform two-phase coexistence MD simulations with LAMMPS \citep{thompson2022a}. To this end, we first equilibrate simulations of pure ammonia and hydrogen in the NPT ensemble at required pressure-temperature conditions. We then initiate phase coexistence simulations by joining the two equilibrated phases, as shown in the leftmost panel of Figure \ref{fig:setup}. This system is then allowed to equilibrate over a period of 1-5 ns until it reaches thermal and compositional dynamic equilibrium. These simulations are run using timesteps of 0.5 fs with a Nos\'e-Hoover thermostat and barostat \citep{nose1984a,hoover1985a,martyna1994a,shinoda2004a} with damping times of 10 fs and 100 fs, respectively. To check system-size effects, we performed coexistence simulations with 256 to 131,072 atoms at 1000 K and 10 GPa, and found that the compositions of the coexisting phases are converged for systems of about 8,000 atoms or more (Figure \ref{fig:parity}, bottom right). Our typical coexistence simulation is thus performed at either 16,384 or 20,000 atoms at $x_{\text{H}_2}$=0.67 or 0.33, respectively.

\subsection{Analysis of simulations}

\textbf{Distinguishing phases}. We classify each simulation from its mass- and number-density profiles along the long axis of the cell, over the last quarter of the trajectory or the last 0.5 ns, whichever is shorter. For each configuration we fit these profiles with a theoretically-motivated hyperbolic-tangent function that two coexisting fluids are expected to follow \citep[Figure~\ref{fig:setup}; ][]{cahn1958a,widom1982a}. A stable two-phase system should keep a statistically similar profile from one configuration to the next, whereas in a single fluid the profile would fluctuate randomly. We thus require a correlation of at least 0.8 between successive snapshots from the trajectory. We also require the density difference between the two slabs to exceed twice the fluctuations of the density within them, so that it reflects two distinct phases rather than thermal fluctuations of one fluid. A simulation is two-phase if its profile persists in this way and the density difference exceeds this threshold in at least 90\% of the configurations, and one-phase if the density difference exceeds it in fewer than half of them. In the remaining cases we use the Binder cumulant $U$ \citep{binder1981a}, which measures whether the distribution of the coarse-grained density is unimodal, as for one phase, or bimodal, as for two. $|U|\le0.05$ indicates one phase and $U>0.30$ two phases. Simulations that remain undecided are not used for further analysis. We also test whether the density difference still drifts over the window, from its slope against time corrected for autocorrelation \citep{sokal1997a}.

\textbf{Phase compositions}. To measure the composition of each coexisting phase, we must first decide which atoms belong to which phase. Because the interface between the phases is not flat but fluctuates, we locate it in each sampled configuration by identifying the Gibbs dividing surface \citep[GDS; ][]{willard2010a} using a modified implementation based on \texttt{pytim} \citep{sega2018a} and \citet{deng2023b}. The mass density is coarse-grained with a Gaussian of width 2.5 \AA{} on a periodic 1.0-\AA{} mesh, and its isosurface at the density midway between those of the two bulk phases defines this curved GDS, from which each atom is assigned a signed shortest distance $d$. Averaged as a function of $d$, the coarse-grained density follows a hyperbolic tangent centered on this surface, changing from the density of one bulk phase to that of the other over a width $w$ that we fit in each configuration. Atoms in the interfacial region, $|d|<w$, are excluded, and the remaining atoms are assigned to the bulk phase on their side of the surface. In each bulk phase we count the hydrogen and nitrogen atoms, $n_{\rm H}$ and $n_{\rm N}$, and convert them to equivalent numbers of molecules, $n_{\mathrm{NH_3}}=n_{\rm N}$ and $n_{\mathrm{H_2}}=(n_{\rm H}-3n_{\rm N})/2$. The composition of the phase is then $x=\langle n_{\mathrm{H_2}}\rangle/(\langle n_{\mathrm{H_2}}\rangle+\langle n_{\mathrm{NH_3}}\rangle)$, where angle brackets denote averages over the configurations of the final 0.5 ns. This definition depends only on elemental inventories and therefore remains meaningful when molecules dissociate. Because successive configurations are correlated, we estimate the uncertainty of each mean composition with a circular block bootstrap \citep{politis1992a}, a resampling form of the blocking method \citep{flyvbjerg1989a}, using blocks four times the integrated autocorrelation time \citep{sokal1997a}. These uncertainties are propagated into the thermodynamic model. As a composition-sensitive check for phases that differ in composition but have nearly the same mass density, we repeat the construction using the nitrogen number density; the reported compositions are from the mass-density analysis.

\textbf{Speciation}. We determine the fractions of different species in homogeneous fluids with $x_{\text{H}_2}=0.67$ from independent simulations in the canonical ensemble. These are performed at the mean volume of the corresponding constant-pressure simulations with a time step of 0.1 fs for 5-10 ps. We define a set of atoms as bonded if they remain in mutual proximity for a minimum time interval $\tau$ \citep[e.g.,][]{tamblyn2010a,gupta2025a}. We set $\tau=100$ fs, about ten vibrational periods of the H-H and N-H bonds ($\approx$8-$\approx$10 fs), so that only those atoms that persist through multiple cycles of atomic vibrations count as bonded. To this end, we track 18 of the nearest neighbors of each atom for an analysis window of 1 picosecond. The reported fraction of species $i$ is $f_i=n_i/\sum_j n_j$, where $n_i$ is the number of units of species $i$, averaged over this analysis window.

\subsection{Thermodynamic model}

A single asymmetric (subregular) solution model describes all our coexistence (binodal) curves (surface). The molar Gibbs free energy of mixing can be written as \citep[e.g.,][]{thompson1967a,haselton1980a,saxena2012a}
\begin{equation}
G_{\mathrm{mix}}=\left(Ax_1+Bx_2\right)x_1x_2\left(1-\frac{T}{T_0}+\frac{P}{P_0}\right)+RT\sum_{i=1}^{2}x_i\ln x_i,
\label{eq:gs-original}
\end{equation}
where $x_1+x_2=1$, $A$ and $B$ describe the composition asymmetry, and $T_0$ and $P_0$ are temperature and pressure scales. For the present binary, $x\equiv x_1$ is the mole fraction of H$_2$ and $x_2=1-x$. This model can be rearranged by defining $r=B/A$, $u=A(1-T/T_0)$ and $v=A/P_0$. We retain this composition dependence while allowing $r$, $u$ and $v$ to depend on the reduced temperature $t=T/(1000\ {\rm K})$:
\begin{equation}
G_{\mathrm{mix}}=\left[x_1+r(t)x_2\right]x_1x_2\left[u(t)+Pv(t)\right]+RT\sum_{i=1}^{2}x_i\ln x_i.
\label{eq:gs-family}
\end{equation}
The first term is the excess contribution and the second is the ideal contribution. The conventional form, in which $r$ and $v$ are constant and $u$ is linear in temperature, cannot reproduce the evolution of the coexistence curves and critical pressure across our vast temperature-pressure range. We therefore allow $r$, $u$ and $v$ to depend on temperature. Here $P$ is in GPa and $R=8.314\times10^{-3}$ kJ mol$^{-1}$ K$^{-1}$. The parameters $u$ (kJ mol$^{-1}$) and $v$ (kJ mol$^{-1}$ GPa$^{-1}$, equivalent to cm$^3$ mol$^{-1}$) set the scale of the excess Gibbs energy extrapolated to zero pressure and of the excess volume $\partial G_{\rm ex}/\partial P$, respectively, and $r$ sets the asymmetry of the coexistence curves (Figure \ref{fig:coexistence}). The compositions of the two coexisting phases follow from the common tangent to $G_{\mathrm{mix}}(x)$, at which the chemical potentials of both H$_2$ and NH$_3$ are equal in the two phases.

The temperature dependence of $r$, $u$ and $v$ is guided by symbolic regression \citep[e.g.,][]{udrescu2020a} as implemented in \textsc{PySR} package \citep{cranmer2023a}, which searched among polynomials and scored each candidate by how well its common-tangent compositions match the measured coexisting compositions; the best candidates were then refit with the full likelihood described below. We adopt (Figure \ref{fig:thermodynamic-functions})
\begin{align*}
r(t)&=r_0,\\
u(t)&=u_0+u_1t+u_2t^2,\\
v(t)&=v_0+v_1t^{-4}.
\end{align*}
We fit the six coefficients to the compositions of 38 pairs of coexisting phases (Figure \ref{fig:coexistence}). Each composition enters a Gaussian likelihood whose variance is the sum of the squares of its uncertainty and of a fitted common scatter. The scatter, 0.045 in $x$, dominates, since the uncertainties of the individual compositions are much smaller (median 0.001). The 22 one-phase simulations, and one two-phase simulation at 1500 K and 25 GPa whose compositions could not be determined, enter as soft penalties, as for censored data in regression \citep[e.g.,][]{tobin1958a,kelly2007a}: the former are penalized when the model places their bulk composition inside the two-phase region, and the latter when it places it outside. We sample the joint posterior distribution of the six coefficients with \textsc{Nautilus} \citep{lange2023a}, which combines importance nested sampling \citep{skilling2004a,skilling2006a} with neural-network-assisted exploration of the likelihood-constrained parameter space. A physical constraint of $\mathrm{P}_c\ge0$ is applied between 500 to 4000 K. The posterior medians, with 2$\sigma$ intervals in brackets, are $r_0=1.41$ [1.31, 1.56], $u_0=-3.41$ [$-5.02$, $-2.15$], $u_1=19.5$ [16.4, 22.9] and $u_2=-4.30$ [$-6.28$, $-2.78$] kJ mol$^{-1}$, and $v_0=0.19$ [0.14, 0.24] and $v_1=0.66$ [0.53, 0.86] kJ mol$^{-1}$ GPa$^{-1}$; see Figure \ref{fig:thermodynamic-functions}.

The critical point at each temperature satisfies $\partial^2G_{\mathrm{mix}}/\partial x^2=\partial^3G_{\mathrm{mix}}/\partial x^3=0$. Because $r$ is independent of temperature, the critical composition is constant, $x_c\approx0.4$, and the critical pressure is
\begin{equation}
\mathrm{P}_c=\frac{\omega_c RT-u(t)}{v(t)},
\label{eq:pc}
\end{equation}
where $\omega_c=1.57$ is the value of the dimensionless excess parameter $[u(t)+Pv(t)]/(RT)$ at the critical point. The resulting critical curve is shown in Figure \ref{fig:critical}, and the coexistence curves that the model predicts at fixed pressures are shown in Figure \ref{fig:binodals}.

\bibliography{planet_evo}

@article{tejada2026a,
  title={Sub-Neptune Memories. I. Implications of Inefficient Mantle Cooling and Silicate Rain},
  author={Tejada Arevalo, Roberto and Gupta, Akash and Burrows, Adam and Zheng, Donghao and Tang, Yao and Deng, Jie},
  journal={The Astrophysical Journal},
  volume={1001},
  number={2},
  pages={243},
  year={2026},
  publisher={The American Astronomical Society}
}

@ARTICLE{gupta2025a,
       author = {{Gupta}, Akash and {Stixrude}, Lars and {Schlichting}, Hilke E.},
        title = "{The Miscibility of Hydrogen and Water in Planetary Atmospheres and Interiors}",
      journal = {\apjl},
         year = 2025,
        month = apr,
       volume = {982},
       number = {2},
          eid = {L35},
        pages = {L35},
          doi = {10.3847/2041-8213/adb631},
archivePrefix = {arXiv},
       eprint = {2407.04685},
 primaryClass = {astro-ph.EP},
       adsurl = {https://ui.adsabs.harvard.edu/abs/2025ApJ...982L..35G}
}

@INPROCEEDINGS{skilling2004a,
       author = {{Skilling}, John},
        title = "{Nested Sampling}",
    booktitle = {American Institute of Physics Conference Series},
         year = 2004,
       editor = {{Fischer}, Rainer and {Preuss}, Roland and {Toussaint}, Udo Von},
       series = {American Institute of Physics Conference Series},
       volume = {735},
        month = nov,
        pages = {395-405},
          doi = {10.1063/1.1835238},
       adsurl = {https://ui.adsabs.harvard.edu/abs/2004AIPC..735..395S}
}

@article{skilling2006a,
       author = "Skilling, John",
          doi = "10.1214/06-BA127",
     fjournal = "Bayesian Analysis",
      journal = "Bayesian Anal.",
        month = "12",
       number = "4",
        pages = "833--859",
    publisher = "International Society for Bayesian Analysis",
        title = "Nested sampling for general Bayesian computation",
       volume = "1",
         year = "2006"
}

@ARTICLE{lange2023a,
       author = {{Lange}, Johannes U.},
        title = "{nautilus: boosting Bayesian importance nested sampling with deep learning}",
      journal = {\mnras},
         year = 2023,
        month = oct,
       volume = {525},
       number = {2},
        pages = {3181-3194},
          doi = {10.1093/mnras/stad2441},
archivePrefix = {arXiv},
       eprint = {2306.16923},
 primaryClass = {astro-ph.IM}
}

@ARTICLE{saumon1995a,
       author = {{Saumon}, D. and {Chabrier}, G. and {van Horn}, H.~M.},
        title = "{An Equation of State for Low-Mass Stars and Giant Planets}",
      journal = {\apjs},
         year = "1995",
        month = "Aug",
       volume = {99},
        pages = {713},
          doi = {10.1086/192204},
       adsurl = {https://ui.adsabs.harvard.edu/abs/1995ApJS...99..713S}
}

@ARTICLE{bean2021a,
       author = {{Bean}, Jacob L. and {Raymond}, Sean N. and {Owen}, James E.},
        title = "{The Nature and Origins of Sub-Neptune Size Planets}",
      journal = {Journal of Geophysical Research (Planets)},
         year = 2021,
        month = jan,
       volume = {126},
       number = {1},
          eid = {e06639},
        pages = {e06639},
          doi = {10.1029/2020JE006639},
archivePrefix = {arXiv},
       eprint = {2010.11867},
 primaryClass = {astro-ph.EP},
       adsurl = {https://ui.adsabs.harvard.edu/abs/2021JGRE..12606639B}
}

@ARTICLE{kresse1996a,
       author = {{Kresse}, G. and {Furthm{\"u}ller}, J.},
        title = "{Efficient iterative schemes for ab initio total-energy calculations using a plane-wave basis set}",
      journal = {\prb},
         year = 1996,
        month = oct,
       volume = {54},
       number = {16},
        pages = {11169-11186},
          doi = {10.1103/PhysRevB.54.11169},
       adsurl = {https://ui.adsabs.harvard.edu/abs/1996PhRvB..5411169K}
}

@ARTICLE{kresse1999a,
       author = {{Kresse}, G. and {Joubert}, D.},
        title = "{From ultrasoft pseudopotentials to the projector augmented-wave method}",
      journal = {\prb},
         year = 1999,
        month = jan,
       volume = {59},
       number = {3},
        pages = {1758-1775},
          doi = {10.1103/PhysRevB.59.1758},
       adsurl = {https://ui.adsabs.harvard.edu/abs/1999PhRvB..59.1758K}
}

@ARTICLE{kresse1993a,
       author = {{Kresse}, G. and {Hafner}, J.},
        title = "{Ab initio molecular dynamics for liquid metals}",
      journal = {\prb},
         year = 1993,
        month = jan,
       volume = {47},
       number = {1},
        pages = {558-561},
          doi = {10.1103/PhysRevB.47.558},
       adsurl = {https://ui.adsabs.harvard.edu/abs/1993PhRvB..47..558K}
}

@article{kresse1996b,
title = {Efficiency of ab-initio total energy calculations for metals and semiconductors using a plane-wave basis set},
journal = {Computational Materials Science},
volume = {6},
number = {1},
pages = {15-50},
year = {1996},
issn = {0927-0256},
doi = {https://doi.org/10.1016/0927-0256(96)00008-0},
author = {G. Kresse and J. Furthmüller}
}

@ARTICLE{hoover1985a,
       author = {{Hoover}, William G.},
        title = "{Canonical dynamics: Equilibrium phase-space distributions}",
      journal = {\pra},
         year = 1985,
        month = mar,
       volume = {31},
       number = {3},
        pages = {1695-1697},
          doi = {10.1103/PhysRevA.31.1695},
       adsurl = {https://ui.adsabs.harvard.edu/abs/1985PhRvA..31.1695H}
}

@ARTICLE{flyvbjerg1989a,
       author = {{Flyvbjerg}, H. and {Petersen}, H.~G.},
        title = "{Error estimates on averages of correlated data}",
      journal = {\jcp},
         year = 1989,
        month = jul,
       volume = {91},
       number = {1},
        pages = {461-466},
          doi = {10.1063/1.457480},
       adsurl = {https://ui.adsabs.harvard.edu/abs/1989JChPh..91..461F}
}

@article{willard2010a,
        title={Instantaneous liquid interfaces},
        author={Willard, Adam P and Chandler, David},
        journal={The Journal of Physical Chemistry B},
        volume={114},
        number={5},
        pages={1954--1958},
        year={2010},
        publisher={ACS Publications}
}

@article{widom1982a,
  title={Potential-distribution theory and the statistical mechanics of fluids},
  author={Widom, B},
  journal={The Journal of Physical Chemistry},
  volume={86},
  number={6},
  pages={869--872},
  year={1982},
  publisher={ACS Publications}
}

@article{cahn1958a,
  title={Free energy of a nonuniform system. I. Interfacial free energy},
  author={Cahn, John W and Hilliard, John E},
  journal={The Journal of chemical physics},
  volume={28},
  number={2},
  pages={258--267},
  year={1958},
  publisher={American Institute of Physics}
}

@article{seward1981a,
author = {Seward, T. M. and Franck, E. U.},
title = {The system hydrogen - water up to 440°C and 2500 bar pressure},
journal = {Berichte der Bunsengesellschaft für physikalische Chemie},
volume = {85},
number = {1},
pages = {2-7},
doi = {https://doi.org/10.1002/bbpc.19810850103},
year = {1981}
}

@ARTICLE{bali2013a,
       author = {{Bali}, Enik{\H{o}} and {Aud{\'e}tat}, Andreas and {Keppler}, Hans},
        title = "{Water and hydrogen are immiscible in Earth's mantle}",
      journal = {\nat},
         year = 2013,
        month = mar,
       volume = {495},
       number = {7440},
        pages = {220-222},
          doi = {10.1038/nature11908},
       adsurl = {https://ui.adsabs.harvard.edu/abs/2013Natur.495..220B}
}

@ARTICLE{soubiran2015a,
       author = {{Soubiran}, Fran{\c{c}}ois and {Militzer}, Burkhard},
        title = "{Miscibility Calculations for Water and Hydrogen in Giant Planets}",
      journal = {\apj},
         year = 2015,
        month = jun,
       volume = {806},
       number = {2},
          eid = {228},
        pages = {228},
          doi = {10.1088/0004-637X/806/2/228},
archivePrefix = {arXiv},
       eprint = {1505.07885},
 primaryClass = {astro-ph.EP},
       adsurl = {https://ui.adsabs.harvard.edu/abs/2015ApJ...806..228S}
}

@ARTICLE{nettelmann2013a,
       author = {{Nettelmann}, N. and {Helled}, R. and {Fortney}, J.~J. and {Redmer}, R.},
        title = "{New indication for a dichotomy in the interior structure of Uranus and Neptune from the application of modified shape and rotation data}",
      journal = {\planss},
         year = 2013,
        month = mar,
       volume = {77},
        pages = {143-151},
          doi = {10.1016/j.pss.2012.06.019},
archivePrefix = {arXiv},
       eprint = {1207.2309},
 primaryClass = {astro-ph.EP},
       adsurl = {https://ui.adsabs.harvard.edu/abs/2013P&SS...77..143N}
}

@article{mcwilliams2016a,
  title = {Optical Properties of Fluid Hydrogen at the Transition to a Conducting State},
  author = {McWilliams, R. Stewart and Dalton, D. Allen and Mahmood, Mohammad F. and Goncharov, Alexander F.},
  journal = {Phys. Rev. Lett.},
  volume = {116},
  issue = {25},
  pages = {255501},
  numpages = {6},
  year = {2016},
  month = {Jun},
  publisher = {American Physical Society},
  doi = {10.1103/PhysRevLett.116.255501},
}

@ARTICLE{penny2019a,
       author = {{Penny}, Matthew T. and {Gaudi}, B. Scott and {Kerins}, Eamonn and {Rattenbury}, Nicholas J. and {Mao}, Shude and {Robin}, Annie C. and {Calchi Novati}, Sebastiano},
        title = "{Predictions of the WFIRST Microlensing Survey. I. Bound Planet Detection Rates}",
      journal = {\apjs},
         year = 2019,
        month = mar,
       volume = {241},
       number = {1},
          eid = {3},
        pages = {3},
          doi = {10.3847/1538-4365/aafb69},
archivePrefix = {arXiv},
       eprint = {1808.02490},
 primaryClass = {astro-ph.EP},
       adsurl = {https://ui.adsabs.harvard.edu/abs/2019ApJS..241....3P}
}

@ARTICLE{stanley2004a,
       author = {{Stanley}, Sabine and {Bloxham}, Jeremy},
        title = "{Convective-region geometry as the cause of Uranus' and Neptune's unusual magnetic fields}",
      journal = {\nat},
         year = 2004,
        month = mar,
       volume = {428},
       number = {6979},
        pages = {151-153},
          doi = {10.1038/nature02376},
       adsurl = {https://ui.adsabs.harvard.edu/abs/2004Natur.428..151S}
}

@ARTICLE{bailey2021a,
       author = {{Bailey}, Elizabeth and {Stevenson}, David J.},
        title = "{Thermodynamically Governed Interior Models of Uranus and Neptune}",
      journal = {\psj},
         year = 2021,
        month = apr,
       volume = {2},
       number = {2},
          eid = {64},
        pages = {64},
          doi = {10.3847/PSJ/abd1e0},
archivePrefix = {arXiv},
       eprint = {2012.04166},
 primaryClass = {astro-ph.EP},
       adsurl = {https://ui.adsabs.harvard.edu/abs/2021PSJ.....2...64B}
}

@ARTICLE{scheibe2021a,
       author = {{Scheibe}, Ludwig and {Nettelmann}, Nadine and {Redmer}, Ronald},
        title = "{Thermal evolution of Uranus and Neptune. II. Deep thermal boundary layer}",
      journal = {\aap},
         year = 2021,
        month = jun,
       volume = {650},
          eid = {A200},
        pages = {A200},
          doi = {10.1051/0004-6361/202140663},
archivePrefix = {arXiv},
       eprint = {2105.01359},
 primaryClass = {astro-ph.EP},
       adsurl = {https://ui.adsabs.harvard.edu/abs/2021A&A...650A.200S}
}

@ARTICLE{stevenson1977a,
       author = {{Stevenson}, D.~J. and {Salpeter}, E.~E.},
        title = "{The phase diagram and transport properties for hydrogen-helium fluid planets.}",
      journal = {\apjs},
         year = 1977,
        month = oct,
       volume = {35},
        pages = {221-237},
          doi = {10.1086/190478},
       adsurl = {https://ui.adsabs.harvard.edu/abs/1977ApJS...35..221S}
}

@ARTICLE{stevenson1977b,
       author = {{Stevenson}, D.~J. and {Salpeter}, E.~E.},
        title = "{The dynamics and helium distribution in hydrogen-helium fluid planets.}",
      journal = {\apjs},
         year = 1977,
        month = oct,
       volume = {35},
        pages = {239-261},
          doi = {10.1086/190479},
       adsurl = {https://ui.adsabs.harvard.edu/abs/1977ApJS...35..239S}
}

@ARTICLE{madhusudhan2023a,
       author = {{Madhusudhan}, Nikku and {Sarkar}, Subhajit and {Constantinou}, Savvas and {Holmberg}, M{\r{a}}ns and {Piette}, Anjali A.~A. and {Moses}, Julianne I.},
        title = "{Carbon-bearing Molecules in a Possible Hycean Atmosphere}",
      journal = {\apjl},
         year = 2023,
        month = oct,
       volume = {956},
       number = {1},
          eid = {L13},
        pages = {L13},
          doi = {10.3847/2041-8213/acf577},
archivePrefix = {arXiv},
       eprint = {2309.05566},
 primaryClass = {astro-ph.EP},
       adsurl = {https://ui.adsabs.harvard.edu/abs/2023ApJ...956L..13M}
}

@ARTICLE{pearl1990a,
       author = {{Pearl}, J.~C. and {Conrath}, B.~J. and {Hanel}, R.~A. and {Pirraglia}, J.~A. and {Coustenis}, A.},
        title = "{The albedo, effective temperature, and energy balance of Uranus, as determined from Voyager IRIS data}",
      journal = {\icarus},
         year = 1990,
        month = mar,
       volume = {84},
       number = {1},
        pages = {12-28},
          doi = {10.1016/0019-1035(90)90155-3},
       adsurl = {https://ui.adsabs.harvard.edu/abs/1990Icar...84...12P}
}

@ARTICLE{pearl1991a,
       author = {{Pearl}, J.~C. and {Conrath}, B.~J.},
        title = "{The albedo, effective temperature, and energy balance of Neptune, as determined from Voyager data}",
      journal = {\jgr},
         year = 1991,
        month = oct,
       volume = {96},
        pages = {18921-18930},
          doi = {10.1029/91JA01087},
       adsurl = {https://ui.adsabs.harvard.edu/abs/1991JGR....9618921P}
}

@ARTICLE{leconte2012a,
       author = {{Leconte}, J. and {Chabrier}, G.},
        title = "{A new vision of giant planet interiors: Impact of double diffusive convection}",
      journal = {\aap},
         year = 2012,
        month = apr,
       volume = {540},
          eid = {A20},
        pages = {A20},
          doi = {10.1051/0004-6361/201117595},
archivePrefix = {arXiv},
       eprint = {1201.4483},
 primaryClass = {astro-ph.EP},
       adsurl = {https://ui.adsabs.harvard.edu/abs/2012A&A...540A..20L}
}

@article{haselton1980a,
  title={Thermodynamics of pyrope-grossular garnets and their stabilities at high temperatures and high pressures},
  author={Haselton, HT and Newton, RC},
  journal={Journal of Geophysical Research: Solid Earth},
  volume={85},
  number={B12},
  pages={6973--6982},
  year={1980},
  publisher={Wiley Online Library}
}

@article{thompson1967a,
  title={Thermodynamic properties of simple solutions.},
  author={Thompson Jr, James B},
  journal={Researches in geochemistry},
  pages={340--361},
  year={1967},
  publisher={John Wiley}
}

@ARTICLE{saxena2012a,
        author = {{Saxena}, Surendra K.},
        title = "{Thermodynamics of Rock-Forming Crystalline Solutions}",
        publisher = {Springer Berlin},
        year = 2012,
        doi = {10.1007/978-3-642-65558-6},
}

@ARTICLE{benneke2024a,
       author = {{Benneke}, Bj{\"o}rn and {Roy}, Pierre-Alexis and {Coulombe}, Louis-Philippe and {Radica}, Michael and {Piaulet}, Caroline and {Ahrer}, Eva-Maria and {Pierrehumbert}, Raymond and {Krissansen-Totton}, Joshua and {Schlichting}, Hilke E. and {Hu}, Renyu and {Yang}, Jeehyun and {Christie}, Duncan and {Thorngren}, Daniel and {Young}, Edward D. and {Pelletier}, Stefan and {Knutson}, Heather A. and {Miguel}, Yamila and {Evans-Soma}, Thomas M. and {Dorn}, Caroline and {Gagnebin}, Anna and {Fortney}, Jonathan J. and {Komacek}, Thaddeus and {MacDonald}, Ryan and {Raul}, Eshan and {Cloutier}, Ryan and {Acuna}, Lorena and {Lafreni{\`e}re}, David and {Cadieux}, Charles and {Doyon}, Ren{\'e} and {Welbanks}, Luis and {Allart}, Romain},
        title = "{JWST Reveals CH$_4$, CO$_2$, and H$_2$O in a Metal-rich Miscible Atmosphere on a Two-Earth-Radius Exoplanet}",
      journal = {arXiv e-prints},
         year = 2024,
        month = mar,
          eid = {arXiv:2403.03325},
        pages = {arXiv:2403.03325},
          doi = {10.48550/arXiv.2403.03325},
archivePrefix = {arXiv},
       eprint = {2403.03325},
 primaryClass = {astro-ph.EP},
       adsurl = {https://ui.adsabs.harvard.edu/abs/2024arXiv240303325B}
}

@ARTICLE{scheibe2019a,
       author = {{Scheibe}, Ludwig and {Nettelmann}, Nadine and {Redmer}, Ronald},
        title = "{Thermal evolution of Uranus and Neptune. I. Adiabatic models}",
      journal = {\aap},
         year = 2019,
        month = dec,
       volume = {632},
          eid = {A70},
        pages = {A70},
          doi = {10.1051/0004-6361/201936378},
archivePrefix = {arXiv},
       eprint = {1911.00447},
 primaryClass = {astro-ph.EP},
       adsurl = {https://ui.adsabs.harvard.edu/abs/2019A&A...632A..70S}
}

@article{mermin1965a,
  title = {Thermal Properties of the Inhomogeneous Electron Gas},
  author = {Mermin, N. David},
  journal = {Phys. Rev.},
  volume = {137},
  issue = {5A},
  pages = {A1441--A1443},
  numpages = {0},
  year = {1965},
  month = {Mar},
  publisher = {American Physical Society},
  doi = {10.1103/PhysRev.137.A1441},
  url = {https://link.aps.org/doi/10.1103/PhysRev.137.A1441}
}

@ARTICLE{tamblyn2010a,
       author = {{Tamblyn}, Isaac and {Bonev}, Stanimir A.},
        title = "{Structure and Phase Boundaries of Compressed Liquid Hydrogen}",
      journal = {\prl},
         year = 2010,
        month = feb,
       volume = {104},
       number = {6},
          eid = {065702},
        pages = {065702},
          doi = {10.1103/PhysRevLett.104.065702},
archivePrefix = {arXiv},
       eprint = {0910.1798},
 primaryClass = {cond-mat.mtrl-sci},
       adsurl = {https://ui.adsabs.harvard.edu/abs/2010PhRvL.104f5702T}
}

@ARTICLE{asplund2009a,
       author = {{Asplund}, Martin and {Grevesse}, Nicolas and {Sauval}, A. Jacques and {Scott}, Pat},
        title = "{The Chemical Composition of the Sun}",
      journal = {\araa},
         year = 2009,
        month = sep,
       volume = {47},
       number = {1},
        pages = {481-522},
          doi = {10.1146/annurev.astro.46.060407.145222},
archivePrefix = {arXiv},
       eprint = {0909.0948},
 primaryClass = {astro-ph.SR},
       adsurl = {https://ui.adsabs.harvard.edu/abs/2009ARA&A..47..481A}
}

@article{nose1984a,
    author = {Nos{\'e}, Shuichi},
    title = {A unified formulation of the constant temperature molecular dynamics methods},
    journal = {The Journal of Chemical Physics},
    volume = {81},
    number = {1},
    pages = {511-519},
    year = {1984},
    month = {07},
    issn = {0021-9606},
    doi = {10.1063/1.447334},
    url = {https://doi.org/10.1063/1.447334},
    eprint = {https://pubs.aip.org/aip/jcp/article-pdf/81/1/511/18949062/511\_1\_online.pdf},
}

@article{zhang2021a,
  title = {Phase Diagram of a Deep Potential Water Model},
  author = {Zhang, Linfeng and Wang, Han and Car, Roberto and E, Weinan},
  journal = {Phys. Rev. Lett.},
  volume = {126},
  issue = {23},
  pages = {236001},
  numpages = {7},
  year = {2021},
  month = {Jun},
  publisher = {American Physical Society},
  doi = {10.1103/PhysRevLett.126.236001},
  url = {https://link.aps.org/doi/10.1103/PhysRevLett.126.236001}
}

@ARTICLE{udrescu2020a,
       author = {{Udrescu}, Silviu-Marian and {Tegmark}, Max},
        title = "{AI Feynman: A physics-inspired method for symbolic regression}",
      journal = {Science Advances},
         year = 2020,
        month = apr,
       volume = {6},
       number = {16},
        pages = {eaay2631},
          doi = {10.1126/sciadv.aay2631},
archivePrefix = {arXiv},
       eprint = {1905.11481},
 primaryClass = {physics.comp-ph},
       adsurl = {https://ui.adsabs.harvard.edu/abs/2020SciA....6.2631U}
}

@ARTICLE{sega2018a,
       author = {{Sega}, Marcello and {Hantal}, Gy{\"o}rgy and {F{\'a}bi{\'a}n}, Bal{\'a}zs and {Jedlovszky}, P{\'a}l},
        title = "{Pytim: A python package for the interfacial analysis of molecular simulations}",
      journal = {Journal of Computational Chemistry},
         year = 2018,
        month = sep,
       volume = {39},
       number = {25},
        pages = {2118-2125},
          doi = {10.1002/jcc.25384}
}

@article{furness2020a,
  title={Accurate and numerically efficient r2SCAN meta-generalized gradient approximation},
  author={Furness, James W and Kaplan, Aaron D and Ning, Jinliang and Perdew, John P and Sun, Jianwei},
  journal={The journal of physical chemistry letters},
  volume={11},
  number={19},
  pages={8208--8215},
  year={2020},
  publisher={ACS Publications}
}

@article{wang2018a,
  title={DeePMD-kit: A deep learning package for many-body potential energy representation and molecular dynamics},
  author={Wang, Han and Zhang, Linfeng and Han, Jiequn and others},
  journal={Computer Physics Communications},
  volume={228},
  pages={178--184},
  year={2018},
  publisher={Elsevier}
}

@article{li2017a,
  author  = {Li, C. and Ingersoll, A. and Janssen, M. and others},
  title   = {The distribution of ammonia on {Jupiter} from a preliminary inversion of {Juno} microwave radiometer data},
  journal = {Geophys. Res. Lett.}, volume = {44}, number = {11}, pages = {5317--5325}, year = {2017},
  doi     = {10.1002/2017GL073159}}

@article{guillot2020a,
  author  = {Guillot, T. and Stevenson, D. J. and Atreya, S. K. and Bolton, S. J. and Becker, H. N.},
  title   = {Storms and the Depletion of Ammonia in {Jupiter}: {I}. Microphysics of ``Mushballs''},
  journal = {J. Geophys. Res. Planets}, volume = {125}, pages = {e2020JE006403}, year = {2020},
  doi     = {10.1029/2020JE006403}}

@article{ohno2023a,
  author  = {Ohno, Kazumasa and Fortney, Jonathan J.},
  title   = {Nitrogen as a Tracer of Giant Planet Formation. {I}. A Universal Deep Adiabatic Profile and Semi-analytical Predictions of Disequilibrium Ammonia Abundances in Warm Exoplanetary Atmospheres},
  journal = {Astrophys. J.}, volume = {946}, number = {1}, pages = {18}, year = {2023},
  doi     = {10.3847/1538-4357/acafed}}

@article{ohno2023b,
  author  = {Ohno, Kazumasa and Fortney, Jonathan J.},
  title   = {Nitrogen as a Tracer of Giant Planet Formation. {II}. Comprehensive Study of Nitrogen Photochemistry and Implications for Observing {NH$_3$} and {HCN} in Transmission and Emission Spectra},
  journal = {Astrophys. J.}, volume = {956}, number = {2}, pages = {125}, year = {2023},
  doi     = {10.3847/1538-4357/ace531}}

@article{hu2021a,
  author  = {Hu, Renyu},
  title   = {Photochemistry and Spectral Characterization of Temperate and Gas-rich Exoplanets},
  journal = {Astrophys. J.}, volume = {921}, number = {1}, pages = {27}, year = {2021},
  doi     = {10.3847/1538-4357/ac1789}}

@article{hu2021b,
  author  = {Hu, Renyu and Damiano, Mario and Scheucher, Markus and Kite, Edwin and Seager, Sara and Rauer, Heike},
  title   = {Unveiling Shrouded Oceans on Temperate Sub-Neptunes via Transit Signatures of Solubility Equilibria versus Gas Thermochemistry},
  journal = {Astrophys. J. Lett.}, volume = {921}, number = {1}, pages = {L8}, year = {2021},
  doi     = {10.3847/2041-8213/ac1f92}}

@article{cavazzoni1999a,
  author  = {Cavazzoni, C. and Chiarotti, G. L. and Scandolo, S. and Tosatti, E. and Bernasconi, M. and Parrinello, M.},
  title   = {Superionic and Metallic States of Water and Ammonia at Giant Planet Conditions},
  journal = {Science}, volume = {283}, number = {5398}, pages = {44--46}, year = {1999},
  doi     = {10.1126/science.283.5398.44}}

@article{chidester2011a,
  author  = {Chidester, B. A. and Strobel, T. A.},
  title   = {The Ammonia--Hydrogen System under Pressure},
  journal = {J. Phys. Chem. A}, volume = {115}, number = {38}, pages = {10433--10437}, year = {2011},
  doi     = {10.1021/jp202779f}}

@article{wiebe1934a,
  author  = {Wiebe, R. and Tremearne, T. H.},
  title   = {The Solubility of Hydrogen in Liquid Ammonia at 25, 50, 75 and 100$^\circ$ and at Pressures to 1000 Atmospheres},
  journal = {J. Am. Chem. Soc.}, volume = {56}, number = {11}, pages = {2357--2360}, year = {1934},
  doi     = {10.1021/ja01326a039}}

@article{michels1950a,
  author  = {Michels, A. and Skelton, G. F. and Dumoulin, E.},
  title   = {Gas--liquid phase equilibrium in the system ammonia--hydrogen--nitrogen},
  journal = {Physica}, volume = {16}, number = {11--12}, pages = {831--838}, year = {1950},
  doi     = {10.1016/0031-8914(50)90092-9}}

@article{militzer2024a,
  author  = {Militzer, B.},
  title   = {Phase separation of planetary ices explains nondipolar magnetic fields of {Uranus} and {Neptune}},
  journal = {Proc. Natl. Acad. Sci. USA}, volume = {121}, number = {49}, pages = {e2403981121}, year = {2024},
  doi     = {10.1073/pnas.2403981121}}

@article{morales2009a,
  author  = {Morales, M. A. and Schwegler, E. and Ceperley, D. and Pierleoni, C. and Hamel, S. and Caspersen, K.},
  title   = {Phase separation in hydrogen--helium mixtures at {Mbar} pressures},
  journal = {Proc. Natl. Acad. Sci. USA}, volume = {106}, number = {5}, pages = {1324--1329}, year = {2009},
  doi     = {10.1073/pnas.0812581106}}

@article{schottler2018a,
  author  = {Sch{\"o}ttler, M. and Redmer, R.},
  title   = {Ab Initio Calculation of the Miscibility Diagram for Hydrogen--Helium Mixtures},
  journal = {Phys. Rev. Lett.}, volume = {120}, pages = {115703}, year = {2018},
  doi     = {10.1103/PhysRevLett.120.115703}}

@article{bergermann2024a,
  author  = {Bergermann, A. and French, M. and Redmer, R.},
  title   = {Ab initio calculation of the miscibility diagram for mixtures of hydrogen and water},
  journal = {Phys. Rev. B}, volume = {109}, pages = {174107}, year = {2024},
  doi     = {10.1103/PhysRevB.109.174107}}

@article{howard2025a,
  author  = {Howard, S. and Helled, R. and Bergermann, A. and Redmer, R.},
  title   = {The possibility of hydrogen--water demixing in {Uranus}, {Neptune}, {K2-18\,b} and {TOI-270\,d}},
  journal = {Astron. Astrophys.}, volume = {703}, pages = {A154}, year = {2025},
  doi     = {10.48550/arXiv.2507.06288}}

@article{canoamoros2024a,
  author  = {Cano Amoros, M. and Nettelmann, N. and Tosi, N. and Baumeister, P. and Rauer, H.},
  title   = {{H$_2$--H$_2$O} demixing in {Uranus} and {Neptune}: Adiabatic structure models},
  journal = {Astron. Astrophys.}, volume = {692}, pages = {A152}, year = {2024},
  doi     = {10.1051/0004-6361/202452148}}

@article{behler2007a,
  author  = {Behler, J. and Parrinello, M.},
  title   = {Generalized Neural-Network Representation of High-Dimensional Potential-Energy Surfaces},
  journal = {Phys. Rev. Lett.}, volume = {98}, pages = {146401}, year = {2007},
  doi     = {10.1103/PhysRevLett.98.146401}}

@article{zhang2018a,
  author  = {Zhang, L. and Han, J. and Wang, H. and Car, R. and E, W.},
  title   = {Deep Potential Molecular Dynamics: A Scalable Model with the Accuracy of Quantum Mechanics},
  journal = {Phys. Rev. Lett.}, volume = {120}, pages = {143001}, year = {2018},
  doi     = {10.1103/PhysRevLett.120.143001}}

@article{bartok2013a,
  author  = {Bart{\'o}k, Albert P. and Kondor, Risi and Cs{\'a}nyi, G{\'a}bor},
  title   = {On Representing Chemical Environments},
  journal = {Phys. Rev. B}, volume = {87}, number = {18}, pages = {184115}, year = {2013},
  doi     = {10.1103/PhysRevB.87.184115}}

@article{cheng2020a,
  author  = {Cheng, Bingqing and Griffiths, Ryan-Rhys and Wengert, Simon and Kunkel, Christian and Stenczel, Tamas and Zhu, Bonan and Deringer, Volker L. and Bernstein, Noam and Margraf, Johannes T. and Reuter, Karsten and Csanyi, Gabor},
  title   = {Mapping Materials and Molecules},
  journal = {Acc. Chem. Res.}, volume = {53}, number = {9}, pages = {1981--1991}, year = {2020},
  doi     = {10.1021/acs.accounts.0c00403}}

@article{tribello2014a,
  author  = {Tribello, G. A. and Bonomi, M. and Branduardi, D. and Camilloni, C. and Bussi, G.},
  title   = {{PLUMED} 2: New feathers for an old bird},
  journal = {Comput. Phys. Commun.}, volume = {185}, number = {2}, pages = {604--613}, year = {2014},
  doi     = {10.1016/j.cpc.2013.09.018}}

@ARTICLE{rogers2025a,
  author = {{Rogers}, James G and {Young}, Edward D and {Schlichting}, Hilke E},
  title = {{Redefining interiors and envelopes: hydrogen--silicate miscibility and its consequences for the structure and evolution of sub-Neptunes}},
  journal = {Monthly Notices of the Royal Astronomical Society},
  year = 2025,
  volume = {544},
  pages = {3496-3511},
  doi = {10.1093/mnras/staf1940},
  eprint = {2509.13320},
}

@ARTICLE{wiebe1937a,
       author = {{Wiebe}, R. and {Gaddy}, V.~L.},
        title = "{The Solubility in Liquid Ammonia of Hydrogen at 0{\textdegree} and of Nitrogen at 0, 50, 75, 90 and 100{\textdegree} at Pressures to 1000 Atmospheres. Critical Phenomena of Ammonia-Nitrogen Mixtures}",
      journal = {Journal of the American Chemical Society},
         year = 1937,
       volume = {59},
       number = {10},
        pages = {1984-1987},
          doi = {10.1021/ja01289a055}
}

@ARTICLE{kaminishi1962a,
       author = {{Kaminishi}, G. and {Toriumi}, T.},
        title = "{The concentration of ammonia vapor in compressed nitrogen and hydrogen over liquid ammonia}",
      journal = {Bulletin of the Chemical Research Institute of Non-Aqueous Solutions, Tohoku University},
         year = 1962,
       volume = {11},
        pages = {1-12},
         note = {Bibliographic data from Mace \& Vrabec (2024); not verified against the original}
}

@ARTICLE{krichevskii1939a,
       author = {{Krichevskii}, I. and {Khazanova}, N.},
        title = "{Content of ammonia in compressed hydrogen and nitrogen in equilibrium with liquid ammonia}",
      journal = {Zhurnal Fizicheskoi Khimii},
         year = 1939,
       volume = {13},
        pages = {1690-1693},
         note = {Bibliographic data from Mace \& Vrabec (2024); not verified against the original}
}

@ARTICLE{ahrens1981a,
       author = {{Ahrens}, M. and {Heusler}, K.~E.},
        title = "{Solubilities of Some Gases in Liquid Ammonia}",
      journal = {Zeitschrift f{\"u}r Physikalische Chemie},
         year = 1981,
       volume = {125},
       number = {1},
        pages = {127-128},
          doi = {10.1524/zpch.1981.125.1.127}
}

@ARTICLE{moore1972a,
       author = {{Moore}, R.~G. and {Otto}, F.~D.},
        title = "{The solubility of hydrogen in aliphatic amines}",
      journal = {The Canadian Journal of Chemical Engineering},
         year = 1972,
       volume = {50},
       number = {3},
        pages = {355-360},
          doi = {10.1002/cjce.5450500308}
}

@ARTICLE{bareli1961a,
       author = {{Bar-Eli}, Kedma and {Klein}, Fritz S.},
        title = "{A Kinetic Method for Measuring Solubilities of Gases in Liquids}",
      journal = {\jcp},
         year = 1961,
       volume = {35},
       number = {5},
        pages = {1915},
          doi = {10.1063/1.1732185}
}

@ARTICLE{thompson2022a,
  author = {Thompson, Aidan P. and Aktulga, H. Metin and Berger, Richard and Bolintineanu, Dan S. and Brown, W. Michael and Crozier, Paul S. and in 't Veld, Pieter J. and Kohlmeyer, Axel and Moore, Stan G. and Nguyen, Trung Dac and others},
  title = {{LAMMPS -- a flexible simulation tool for particle-based materials modeling at the atomic, meso, and continuum scales}},
  journal = {Computer Physics Communications},
  year = 2022,
  volume = {271},
  pages = {108171},
  doi = {10.1016/j.cpc.2021.108171},
}

@ARTICLE{hunter2007a,
  author = {{Hunter}, John D.},
  title = {{Matplotlib: A 2D Graphics Environment}},
  journal = {Computing in Science \& Engineering},
  year = 2007,
  volume = {9},
  pages = {90-95},
  doi = {10.1109/MCSE.2007.55},
}

@ARTICLE{harris2020a,
  author = {{Harris}, Charles R. and {Millman}, K. Jarrod and {van der Walt}, St{\'e}fan J. and {Gommers}, Ralf and {Virtanen}, Pauli and {Cournapeau}, David and {Wieser}, Eric and {Taylor}, Julian and {Berg}, Sebastian and {Smith}, Nathaniel J. and {Kern}, Robert and {Picus}, Matti and {Hoyer}, Stephan and {van Kerkwijk}, Marten H. and {Brett}, Matthew and {Haldane}, Allan and {del R{\'i}o}, Jaime Fern{\'a}ndez and {Wiebe}, Mark and {Peterson}, Pearu and {G{\'e}rard-Marchant}, Pierre and {Sheppard}, Kevin and {Reddy}, Tyler and {Weckesser}, Warren and {Abbasi}, Hameer and {Gohlke}, Christoph and {Oliphant}, Travis E.},
  title = {{Array programming with NumPy}},
  journal = {Nature},
  year = 2020,
  volume = {585},
  pages = {357-362},
  doi = {10.1038/s41586-020-2649-2},
}

@ARTICLE{deng2023b,
  author = {{Deng}, Jie and {Niu}, Haiyang and {Hu}, Junwei and {Chen}, Mingyi and {Stixrude}, Lars},
  title = {{Melting of MgSiO$_3$ determined by machine learning potentials}},
  journal = {Physical Review B},
  year = 2023,
  volume = {107},
  pages = {064103},
  doi = {10.1103/PhysRevB.107.064103},
}

@ARTICLE{gilmore2026a,
  author = {{Gilmore}, Travis and {Stixrude}, Lars},
  title = {{Core--envelope miscibility in sub-Neptunes and super-Earths}},
  journal = {Nature},
  year = 2026,
  volume = {650},
  pages = {60-64},
  doi = {10.1038/s41586-025-09970-4},
}

@article{wogan2024a,
  author  = {Wogan, Nicholas F. and Batalha, Natasha E. and Zahnle, Kevin J. and
             Krissansen-Totton, Joshua and Tsai, Shang-Min and Hu, Renyu},
  title   = {JWST Observations of K2-18b Can Be Explained by a Gas-rich Mini-Neptune with No Habitable Surface},
  journal = {The Astrophysical Journal Letters},
  volume  = {963},
  pages   = {L7},
  year    = {2024},
  doi     = {10.3847/2041-8213/ad2616},
  eprint  = {2401.11082},
  archivePrefix = {arXiv}}

@article{heise1972a,
  author = {Heise, Friedel},
  title = {Fl{\"u}ssigkeits-Dampfgleichgewichte zwischen Ammoniak und verschiedenen Gasen},
  journal = {Berichte der Bunsengesellschaft f{\"u}r physikalische Chemie},
  volume = {76},
  number = {9},
  pages = {938--943},
  year = {1972},
  doi = {10.1002/bbpc.19720760914}}

@article{reamer1959a,
  author = {Reamer, H. H. and Sage, B. H.},
  title = {Phase Behavior in the Hydrogen-Ammonia System},
  journal = {Journal of Chemical and Engineering Data},
  volume = {4},
  number = {2},
  pages = {152--154},
  year = {1959},
  doi = {10.1021/je60002a012}}

@article{cranmer2023a,
  author = {Cranmer, Miles},
  title = {Interpretable Machine Learning for Science with PySR and SymbolicRegression.jl},
  journal = {arXiv e-prints},
  year = {2023},
  eprint = {2305.01582},
  archivePrefix = {arXiv},
  doi = {10.48550/arXiv.2305.01582}}

@article{ninet2012a,
  author = {Ninet, S. and Datchi, F. and Saitta, A. M.},
  title = {Proton Disorder and Superionicity in Hot Dense Ammonia Ice},
  journal = {Physical Review Letters},
  volume = {108},
  number = {16},
  pages = {165702},
  year = {2012},
  doi = {10.1103/PhysRevLett.108.165702}}

@article{bethkenhagen2013a,
  author = {Bethkenhagen, Mandy and French, Martin and Redmer, Ronald},
  title = {Equation of state and phase diagram of ammonia at high pressures from ab initio simulations},
  journal = {The Journal of Chemical Physics},
  volume = {138},
  number = {23},
  pages = {234504},
  year = {2013},
  doi = {10.1063/1.4810883}}

@article{hernandez2023a,
  author = {Hernandez, J.-A. and Bethkenhagen, M. and Ninet, S. and French, M. and Benuzzi-Mounaix, A. and Datchi, F. and Guarguaglini, M. and Lefevre, F. and Occelli, F. and Redmer, R. and Vinci, T. and Ravasio, A.},
  title = {Melting curve of superionic ammonia at planetary interior conditions},
  journal = {Nature Physics},
  volume = {19},
  pages = {1280--1285},
  year = {2023},
  doi = {10.1038/s41567-023-02074-8}}

@ARTICLE{madhusudhan2023b,
       author = {{Madhusudhan}, Nikku and {Moses}, Julianne I. and {Rigby}, Frances and {Barrier}, Edouard},
        title = "{Chemical conditions on Hycean worlds}",
      journal = {Faraday Discussions},
         year = 2023,
        month = sep,
       volume = {245},
        pages = {80-111},
          doi = {10.1039/D3FD00075C},
archivePrefix = {arXiv},
       eprint = {2306.13706},
 primaryClass = {astro-ph.EP},
       adsurl = {https://ui.adsabs.harvard.edu/abs/2023FaDi..245...80M}
}

@ARTICLE{phillips2021a,
       author = {{Phillips}, Caprice L. and {Wang}, Ji and {Kendrew}, Sarah and {Greene}, Thomas P. and {Hu}, Renyu and {Valenti}, Jeff and {Panero}, Wendy R. and {Schulze}, Joseph},
        title = "{Detecting Biosignatures in the Atmospheres of Gas Dwarf Planets with the James Webb Space Telescope}",
      journal = {\apj},
         year = 2021,
        month = dec,
       volume = {923},
       number = {2},
          eid = {144},
        pages = {144},
          doi = {10.3847/1538-4357/ac29be},
archivePrefix = {arXiv},
       eprint = {2109.12132},
 primaryClass = {astro-ph.EP},
       adsurl = {https://ui.adsabs.harvard.edu/abs/2021ApJ...923..144P}
}

@ARTICLE{giles2017a,
       author = {{Giles}, Rohini S. and {Fletcher}, Leigh N. and {Irwin}, Patrick G.~J. and {Orton}, Glenn S. and {Sinclair}, James A.},
        title = "{Ammonia in Jupiter's Troposphere From High-Resolution 5 {\ensuremath{\mu}}m Spectroscopy}",
      journal = {\grl},
         year = 2017,
        month = nov,
       volume = {44},
       number = {21},
        pages = {10,838-10,844},
          doi = {10.1002/2017GL075221},
archivePrefix = {arXiv},
       eprint = {1710.03336},
 primaryClass = {astro-ph.EP},
       adsurl = {https://ui.adsabs.harvard.edu/abs/2017GeoRL..4410838G}
}

@ARTICLE{irwin2025a,
       author = {{Irwin}, Patrick G.~J. and {Hill}, Steven M. and {Fletcher}, Leigh N. and {Alexander}, Charlotte and {Rogers}, John H.},
        title = "{Clouds and Ammonia in the Atmospheres of Jupiter and Saturn Determined From a Band-Depth Analysis of VLT/MUSE Observations}",
      journal = {Journal of Geophysical Research (Planets)},
         year = 2025,
        month = jan,
       volume = {130},
       number = {1},
        pages = {2024JE008622},
          doi = {10.1029/2024JE008622},
       adsurl = {https://ui.adsabs.harvard.edu/abs/2025JGRE..13008622I}
}

@ARTICLE{hubbard1980a,
       author = {{Hubbard}, W.~B. and {Macfarlane}, J.~J.},
        title = "{Structure and evolution of Uranus and Neptune.}",
      journal = {\jgr},
         year = 1980,
        month = jan,
       volume = {85},
        pages = {225-234},
          doi = {10.1029/JB085iB01p00225},
       adsurl = {https://ui.adsabs.harvard.edu/abs/1980JGR....85..225H}
}

@article{horn2025a,
    title = {Building wet planets through high-pressure magma–hydrogen reactions},
    volume = {646},
    issn = {0028-0836, 1476-4687},
    url = {https://www.nature.com/articles/s41586-025-09630-7},
    doi = {10.1038/s41586-025-09630-7},
    language = {en},
    number = {8087},
    urldate = {2025-12-03},
    journal = {Nature},
    author = {Horn, H. W. and Vazan, A. and Chariton, S. and Prakapenka, V. B. and Shim, S.-H.},
    month = oct,
    year = {2025},
    pages = {1069--1074},
}

@ARTICLE{insixiengmay2025a,
       author = {{Insixiengmay}, Leslie and {Stixrude}, Lars},
        title = "{MgO miscibility in liquid iron}",
      journal = {Earth and Planetary Science Letters},
         year = 2025,
        month = mar,
       volume = {654},
          eid = {119242},
        pages = {119242},
          doi = {10.1016/j.epsl.2025.119242},
archivePrefix = {arXiv},
       eprint = {2409.07681},
 primaryClass = {physics.geo-ph},
       adsurl = {https://ui.adsabs.harvard.edu/abs/2025E&PSL.65419242I}
}

@ARTICLE{queyroux2019a,
       author = {{Queyroux}, Jean-Antoine and {Ninet}, Sandra and {Weck}, Gunnar and {Garbarino}, Gaston and {Mezouar}, Mohamed and {Datchi}, Fr{\'e}d{\'e}ric},
        title = "{Structure of liquid ammonia at high pressures and temperatures}",
      journal = {\prb},
         year = 2019,
        month = dec,
       volume = {100},
       number = {22},
          eid = {224104},
        pages = {224104},
          doi = {10.1103/PhysRevB.100.224104},
       adsurl = {https://ui.adsabs.harvard.edu/abs/2019PhRvB.100v4104Q}
}

@ARTICLE{tejada2025a,
       author = {{Tejada Arevalo}, Roberto},
        title = "{Different Inhomogeneous Evolutionary Histories for Uranus and Neptune}",
      journal = {\apjl},
         year = 2025,
        month = aug,
       volume = {989},
       number = {2},
          eid = {L40},
        pages = {L40},
          doi = {10.3847/2041-8213/adf3a5},
archivePrefix = {arXiv},
       eprint = {2506.13857},
 primaryClass = {astro-ph.EP},
       adsurl = {https://ui.adsabs.harvard.edu/abs/2025ApJ...989L..40T}
}

@ARTICLE{rigby2026a,
       author = {{Rigby}, Frances E. and {Madhusudhan}, Nikku},
        title = "{The surface and interior conditions of temperate sub-Neptune TOI-270 d}",
      journal = {\mnras},
         year = 2026,
        month = jan,
       volume = {545},
       number = {1},
          eid = {staf1969},
        pages = {staf1969},
          doi = {10.1093/mnras/staf1969},
archivePrefix = {arXiv},
       eprint = {2511.16722},
 primaryClass = {astro-ph.EP},
       adsurl = {https://ui.adsabs.harvard.edu/abs/2026MNRAS.545f1969R}
}

@ARTICLE{ravasio2021a,
       author = {{Ravasio}, A. and {Bethkenhagen}, M. and {Hernandez}, J.-A. and {Benuzzi-Mounaix}, A. and {Datchi}, F. and {French}, M. and {Guarguaglini}, M. and {Lefevre}, F. and {Ninet}, S. and {Redmer}, R. and {Vinci}, T.},
        title = "{Metallization of Shock-Compressed Liquid Ammonia}",
      journal = {\prl},
         year = 2021,
        month = jan,
       volume = {126},
       number = {2},
          eid = {025003},
        pages = {025003},
          doi = {10.1103/PhysRevLett.126.025003},
archivePrefix = {arXiv},
       eprint = {2101.06692},
 primaryClass = {astro-ph.EP},
       adsurl = {https://ui.adsabs.harvard.edu/abs/2021PhRvL.126b5003R}
}

@ARTICLE{sumi2010a,
       author = {{Sumi}, T. and {Bennett}, D.~P. and {Bond}, I.~A. and {Udalski}, A. and {Batista}, V. and others},
        title = "{A Cold Neptune-Mass Planet OGLE-2007-BLG-368Lb: Cold Neptunes Are Common}",
      journal = {\apj},
         year = 2010,
       volume = {710},
       number = {2},
        pages = {1641},
          doi = {10.1088/0004-637X/710/2/1641}
}

@ARTICLE{muraki2011a,
       author = {{Muraki}, Y. and {Han}, C. and {Bennett}, D.~P. and {Suzuki}, D. and {Monard}, L.~A.~G. and others},
        title = "{Discovery and Mass Measurements of a Cold, 10 Earth Mass Planet and Its Host Star}",
      journal = {\apj},
         year = 2011,
       volume = {741},
       number = {1},
          eid = {22},
        pages = {22},
          doi = {10.1088/0004-637X/741/1/22}
}

@ARTICLE{koshimoto2017a,
       author = {{Koshimoto}, N. and {Udalski}, A. and {Beaulieu}, J.~P. and {Sumi}, T. and {Bennett}, D.~P. and others},
        title = "{OGLE-2012-BLG-0950Lb: The First Planet Mass Measurement from Only Microlens Parallax and Lens Flux}",
      journal = {\aj},
         year = 2017,
       volume = {153},
       number = {1},
          eid = {1},
        pages = {1},
          doi = {10.3847/1538-3881/153/1/1}
}

@ARTICLE{molter2021a,
       author = {{Molter}, Edward M. and {de Pater}, Imke and {Luszcz-Cook}, Statia and {Tollefson}, Joshua and {Sault}, Robert J. and others},
        title = "{Tropospheric Composition and Circulation of Uranus with ALMA and the VLA}",
      journal = {\psj},
         year = 2021,
       volume = {2},
       number = {1},
          eid = {3},
        pages = {3},
          doi = {10.3847/PSJ/abc48a}
}

@ARTICLE{tollefson2021a,
       author = {{Tollefson}, Joshua and {de Pater}, Imke and {Molter}, Edward M. and {Sault}, Robert J. and {Butler}, Bryan J. and {Luszcz-Cook}, Statia and {DeBoer}, David},
        title = "{Neptune's Spatial Brightness Temperature Variations from the VLA and ALMA}",
      journal = {\psj},
         year = 2021,
       volume = {2},
       number = {3},
          eid = {105},
        pages = {105},
          doi = {10.3847/PSJ/abf837}
}

@ARTICLE{atreya2020a,
       author = {{Atreya}, Sushil K. and {Hofstadter}, Mark H. and {In}, Joong Hyun and {Mousis}, Olivier and {Reh}, Kim and {Wong}, Michael H.},
        title = "{Deep Atmosphere Composition, Structure, Origin, and Exploration, with Particular Focus on Critical in situ Science at the Icy Giants}",
      journal = {\ssr},
         year = 2020,
       volume = {216},
       number = {1},
          eid = {18},
        pages = {18},
          doi = {10.1007/s11214-020-0640-8}
}

@ARTICLE{valsson2014a,
       author = {{Valsson}, Omar and {Parrinello}, Michele},
        title = "{Variational Approach to Enhanced Sampling and Free Energy Calculations}",
      journal = {\prl},
         year = 2014,
       volume = {113},
       number = {9},
          eid = {090601},
        pages = {090601},
          doi = {10.1103/PhysRevLett.113.090601}
}

@ARTICLE{piaggi2019a,
       author = {{Piaggi}, Pablo M. and {Parrinello}, Michele},
        title = "{Multithermal-Multibaric Molecular Simulations from a Variational Principle}",
      journal = {\prl},
         year = 2019,
       volume = {122},
       number = {5},
          eid = {050601},
        pages = {050601},
          doi = {10.1103/PhysRevLett.122.050601}
}

@article{Bonomi2019a,
  author  = {{PLUMED consortium}},
  title   = {Promoting transparency and reproducibility in enhanced molecular simulations},
  journal = {Nature Methods},
  year    = {2019},
  volume  = {16},
  number  = {8},
  pages   = {670--673},
  doi     = {10.1038/s41592-019-0506-8},
  url     = {https://doi.org/10.1038/s41592-019-0506-8}
}

@article{Bonomi2009a,
  author  = {Bonomi, M. and Branduardi, D. and Bussi, G. and
             Camilloni, C. and Provasi, D. and Raiteri, P. and
             Donadio, D. and Marinelli, F. and Pietrucci, F. and
             Broglia, R. A. and Parrinello, M.},
  title   = {{PLUMED}: A portable plugin for free-energy
             calculations with molecular dynamics},
  journal = {Computer Physics Communications},
  volume  = {180},
  number  = {10},
  pages   = {1961--1972},
  year    = {2009},
  doi     = {10.1016/j.cpc.2009.05.011},
  url     = {https://doi.org/10.1016/j.cpc.2009.05.011},
  eprint  = {0902.0874},
  archivePrefix = {arXiv}
}

@article{Zeng2023a,
  title  = {{DeePMD-kit v2: A software package for deep potential models}},
  author =   {Jinzhe Zeng and Duo Zhang and Denghui Lu and Pinghui Mo and Zeyu Li
         and Yixiao Chen and Mari{\'a}n Rynik and Li'ang Huang and Ziyao Li and 
         Shaochen Shi and Yingze Wang and Haotian Ye and Ping Tuo and Jiabin
         Yang and Ye Ding and Yifan Li and Davide Tisi and Qiyu Zeng and Han 
         Bao and Yu Xia and Jiameng Huang and Koki Muraoka and Yibo Wang and 
         Junhan Chang and Fengbo Yuan and Sigbj{\o}rn L{\o}land Bore and Chun
         Cai and Yinnian Lin and Bo Wang and Jiayan Xu and Jia-Xin Zhu and 
         Chenxing Luo and Yuzhi Zhang and Rhys E A Goodall and Wenshuo Liang
         and Anurag Kumar Singh and Sikai Yao and Jingchao Zhang and Renata
         Wentzcovitch and Jiequn Han and Jie Liu and Weile Jia and Darrin M
         York and Weinan E and Roberto Car and Linfeng Zhang and Han Wang},
  journal =  {J. Chem. Phys.},
  volume =   159,
  issue =    5,  
  year =    2023,
  pages  =   054801,
  doi =      {10.1063/5.0155600},
}

@Article{Zeng2025a,
  author =   {Jinzhe Zeng and Duo Zhang and Anyang Peng and Xiangyu Zhang and Sensen
             He and Yan Wang and Xinzijian Liu and Hangrui Bi and Yifan Li and Chun
             Cai and Chengqian Zhang and Yiming Du and Jia-Xin Zhu and Pinghui Mo
             and Zhengtao Huang and Qiyu Zeng and Shaochen Shi and Xuejian Qin and
             Zhaoxi Yu and Chenxing Luo and Ye Ding and Yun-Pei Liu and Ruosong Shi
             and Zhenyu Wang and Sigbj{\o}rn L{\o}land Bore and Junhan Chang and
             Zhe Deng and Zhaohan Ding and Siyuan Han and Wanrun Jiang and Guolin
             Ke and Zhaoqing Liu and Denghui Lu and Koki Muraoka and Hananeh Oliaei
             and Anurag Kumar Singh and Haohui Que and Weihong Xu and Zhangmancang
             Xu and Yong-Bin Zhuang and Jiayu Dai and Timothy J. Giese and Weile
             Jia and Ben Xu and Darrin M. York and Linfeng Zhang and Han Wang},
    title =    {{DeePMD-kit v3: A Multiple-Backend Framework for Machine Learning
             Potentials}},
  journal =  {J. Chem. Theory Comput.},
  year =     2025,
  volume =   21,
  number =   9,
  pages =    {4375--4385},
  doi =      {10.1021/acs.jctc.5c00340},
}

@ARTICLE{french2012a,
       author = {{French}, Martin and {Becker}, Andreas and {Lorenzen}, Winfried and {Nettelmann}, Nadine and {Bethkenhagen}, Mandy and {Wicht}, Johannes and {Redmer}, Ronald},
        title = "{Ab Initio Simulations for Material Properties along the Jupiter Adiabat}",
      journal = {\apjs},
         year = 2012,
       volume = {202},
       number = {1},
          eid = {5},
        pages = {5},
          doi = {10.1088/0067-0049/202/1/5}
}

@ARTICLE{french2019a,
       author = {{French}, Martin and {Nettelmann}, Nadine},
        title = "{Viscosity and Prandtl Number of Warm Dense Water as in Ice Giant Planets}",
      journal = {\apj},
         year = 2019,
       volume = {881},
       number = {1},
          eid = {81},
        pages = {81},
          doi = {10.3847/1538-4357/ab2d21}
}

@ARTICLE{grasselli2020a,
       author = {{Grasselli}, Federico and {Stixrude}, Lars and {Baroni}, Stefano},
        title = "{Heat and charge transport in H$_2$O at ice-giant conditions from ab initio molecular dynamics simulations}",
      journal = {Nature Communications},
         year = 2020,
       volume = {11},
          eid = {3605},
        pages = {3605},
          doi = {10.1038/s41467-020-17275-5}
}

@ARTICLE{piaggi2019b,
       author = {{Piaggi}, Pablo M. and {Parrinello}, Michele},
        title = "{Calculation of phase diagrams in the multithermal-multibaric ensemble}",
      journal = {\jcp},
         year = 2019,
       volume = {150},
       number = {24},
          eid = {244119},
        pages = {244119},
          doi = {10.1063/1.5102104}
}

@ARTICLE{kingsbury2022a,
       author = {{Kingsbury}, Ryan and {Gupta}, Ayush S. and {Bartel}, Christopher J. and {Munro}, Jason M. and {Dwaraknath}, Shyam and {Horton}, Matthew and {Persson}, Kristin A.},
        title = "{Performance comparison of r$^2$SCAN and SCAN metaGGA density functionals for solid materials via an automated, high-throughput computational workflow}",
      journal = {Physical Review Materials},
         year = 2022,
       volume = {6},
        pages = {013801},
          doi = {10.1103/PhysRevMaterials.6.013801}
}

@ARTICLE{sun2015a,
       author = {{Sun}, Jianwei and {Ruzsinszky}, Adrienn and {Perdew}, John P.},
        title = "{Strongly Constrained and Appropriately Normed Semilocal Density Functional}",
      journal = {Physical Review Letters},
         year = 2015,
       volume = {115},
        pages = {036402},
          doi = {10.1103/PhysRevLett.115.036402}
}

@ARTICLE{chattopadhyay2026a,
       author = {{Chattopadhyay}, Rajorshi and {Jahn}, Sandro},
        title = "{Vibrational and Structural Properties of Aqueous H$_2$SO$_4$ and Na$_2$SO$_4$ Systems from Ambient to Supercritical Conditions: A Comparative Study between GGA(-D3) and r2SCAN Functionals}",
      journal = {The Journal of Physical Chemistry A},
         year = 2026,
       volume = {130},
       number = {28},
        pages = {5309--5318},
          doi = {10.1021/acs.jpca.6c03132}
}

@INPROCEEDINGS{kingma2015a,
       author = {{Kingma}, Diederik P. and {Ba}, Jimmy},
        title = "{Adam: A Method for Stochastic Optimization}",
    booktitle = {3rd International Conference on Learning Representations},
         year = 2015,
          url = {https://arxiv.org/abs/1412.6980}
}

@ARTICLE{ZhangYuzhi2020a,
       author = {{Zhang}, Yuzhi and {Wang}, Haidi and {Chen}, Weijie and {Zeng}, Jinzhe and {Zhang}, Linfeng and {Wang}, Han and {E}, Weinan},
        title = "{DP-GEN: A concurrent learning platform for the generation of reliable deep learning based potential energy models}",
      journal = {Computer Physics Communications},
         year = 2020,
        month = aug,
       volume = {253},
          eid = {107206},
        pages = {107206},
          doi = {10.1016/j.cpc.2020.107206},
archivePrefix = {arXiv},
       eprint = {1910.12690},
 primaryClass = {physics.comp-ph},
       adsurl = {https://ui.adsabs.harvard.edu/abs/2020CoPhC.25307206Z}
}

@ARTICLE{burrows2006a,
       author = {{Burrows}, Adam and {Sudarsky}, David and {Hubeny}, Ivan},
        title = "{L and T Dwarf Models and the L to T Transition}",
      journal = {\apj},
         year = 2006,
       volume = {640},
       number = {2},
        pages = {1063},
          doi = {10.1086/500293}
}

@BOOK{pierrehumbert2010a,
       author = {{Pierrehumbert}, Raymond T.},
        title = "{Principles of Planetary Climate}",
    publisher = {Cambridge University Press},
      address = {Cambridge},
         year = 2010,
         isbn = {9780521865562},
          doi = {10.1017/CBO9780511780783}
}

@article{shorttle2024a,
  author  = {Shorttle, Oliver and Jordan, Sean and Nicholls, Harrison
             and Lichtenberg, Tim and Bower, Dan J.},
  title   = {Distinguishing Oceans of Water from Magma on Mini-Neptune K2-18b},
  journal = {The Astrophysical Journal Letters},
  year    = {2024},
  volume  = {962},
  number  = {1},
  pages   = {L8},
  doi     = {10.3847/2041-8213/ad206e},
  eprint  = {2401.05864},
  archivePrefix = {arXiv}
}

@article{gao2023a,
  author  = {Gao, Kehui and Wu, Jiangtao and Bell, Ian H.
             and Harvey, Allan H. and Lemmon, Eric W.},
  title   = {A Reference Equation of State with an Associating Term
             for the Thermodynamic Properties of Ammonia},
  journal = {Journal of Physical and Chemical Reference Data},
  year    = {2023},
  volume  = {52},
  number  = {1},
  pages   = {013102},
  doi     = {10.1063/5.0128269}
}

@ARTICLE{binder1981a,
       author = {{Binder}, K.},
        title = "{Finite size scaling analysis of Ising model block distribution functions}",
      journal = {Zeitschrift f{\"u}r Physik B Condensed Matter},
         year = 1981,
       volume = {43},
       number = {2},
        pages = {119-140},
          doi = {10.1007/BF01293604}
}

@INCOLLECTION{sokal1997a,
       author = {{Sokal}, Alan},
        title = "{Monte Carlo Methods in Statistical Mechanics: Foundations and New Algorithms}",
    booktitle = {Functional Integration},
       series = {NATO ASI Series},
    publisher = {Springer},
      address = {Boston, MA},
         year = 1997,
        pages = {131-192},
          doi = {10.1007/978-1-4899-0319-8_6}
}

@INCOLLECTION{politis1992a,
       author = {{Politis}, Dimitris N. and {Romano}, Joseph P.},
        title = "{A Circular Block-Resampling Procedure for Stationary Data}",
    booktitle = {Exploring the Limits of Bootstrap},
       editor = {{LePage}, Raoul and {Billard}, Lynne},
    publisher = {Wiley},
      address = {New York},
         year = 1992,
        pages = {263-270}
}

@INPROCEEDINGS{zhang2018b,
       author = {{Zhang}, Linfeng and {Han}, Jiequn and {Wang}, Han and {Saidi}, Wissam A. and {Car}, Roberto and {E}, Weinan},
        title = "{End-to-end Symmetry Preserving Inter-atomic Potential Energy Model for Finite and Extended Systems}",
    booktitle = {Advances in Neural Information Processing Systems},
         year = 2018,
       editor = {{Bengio}, S. and {Wallach}, H. and {Larochelle}, H. and {Grauman}, K. and {Cesa-Bianchi}, N. and {Garnett}, R.},
       volume = {31},
    publisher = {Curran Associates, Inc.},
        pages = {4436-4446},
          url = {https://proceedings.neurips.cc/paper_files/paper/2018/hash/e2ad76f2326fbc6b56a45a56c59fafdb-Abstract.html},
archivePrefix = {arXiv},
       eprint = {1805.09003},
 primaryClass = {physics.comp-ph}
}

@ARTICLE{blochl1994a,
       author = {{Bl{\"o}chl}, P.~E.},
        title = "{Projector augmented-wave method}",
      journal = {\prb},
         year = 1994,
        month = dec,
       volume = {50},
       number = {24},
        pages = {17953-17979},
          doi = {10.1103/PhysRevB.50.17953}
}

@ARTICLE{martyna1994a,
       author = {{Martyna}, Glenn J. and {Tobias}, Douglas J. and {Klein}, Michael L.},
        title = "{Constant pressure molecular dynamics algorithms}",
      journal = {\jcp},
         year = 1994,
        month = sep,
       volume = {101},
       number = {5},
        pages = {4177-4189},
          doi = {10.1063/1.467468}
}

@ARTICLE{shinoda2004a,
       author = {{Shinoda}, Wataru and {Shiga}, Motoyuki and {Mikami}, Masuhiro},
        title = "{Rapid estimation of elastic constants by molecular dynamics simulation under constant stress}",
      journal = {\prb},
         year = 2004,
        month = apr,
       volume = {69},
       number = {13},
          eid = {134103},
        pages = {134103},
          doi = {10.1103/PhysRevB.69.134103}
}

@ARTICLE{tobin1958a,
       author = {{Tobin}, James},
        title = "{Estimation of Relationships for Limited Dependent Variables}",
      journal = {Econometrica},
         year = 1958,
       volume = {26},
       number = {1},
        pages = {24-36},
          doi = {10.2307/1907382}
}

@ARTICLE{kelly2007a,
       author = {{Kelly}, Brandon C.},
        title = "{Some Aspects of Measurement Error in Linear Regression of Astronomical Data}",
      journal = {\apj},
         year = 2007,
       volume = {665},
       number = {2},
        pages = {1489-1506},
          doi = {10.1086/519947}
}
\bibliographystyle{aasjournal}

\newpage

\section{Supplementary Information}\label{sec:supplementary}
\setcounter{figure}{0}
\renewcommand{\thefigure}{S\arabic{figure}}
\renewcommand{\theHfigure}{S\arabic{figure}}

\begin{figure*}[ht!]
\centering
\includegraphics[width=\textwidth,trim=0 0 0 0,clip]{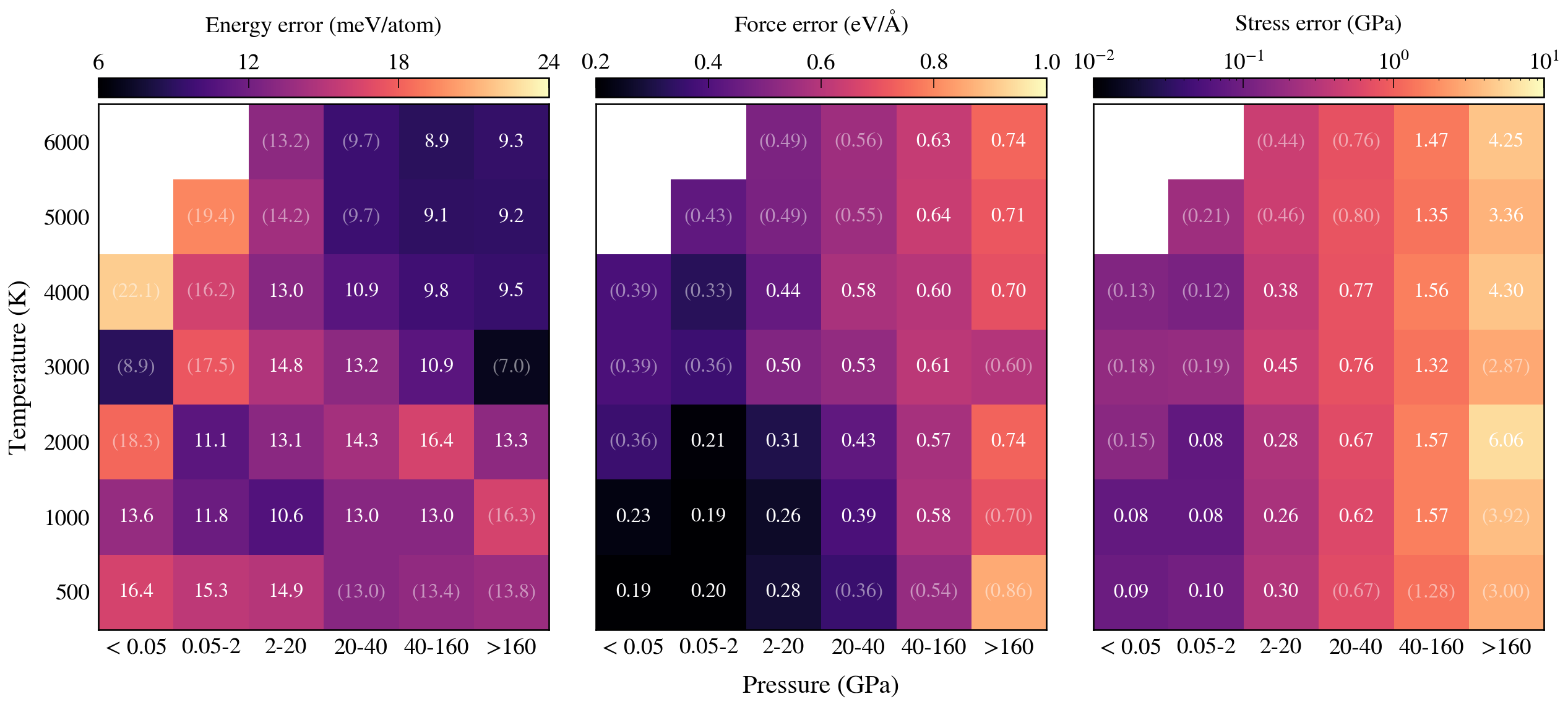}
\caption{Training-set performance of the final machine-learning potential (MLP) across the sampled pressure and temperature space. The panels show the mean absolute energy error per atom and the root-mean-square errors of the force and stress components for 68,994 training configurations, containing 16,128,230 atoms in total. Configurations are binned by the midpoint temperature of their training zone (Methods) and by pressure, which includes the ionic kinetic contribution at that temperature. Values in parentheses mark bins with fewer than 300 configurations, and white cells contain none. The energy and force color bars are linear, and the stress color bar is logarithmic.}
\label{fig:training-errors}
\end{figure*}

\begin{figure*}
\centering
\includegraphics[width=\textwidth,trim=5 0 0 0,clip]{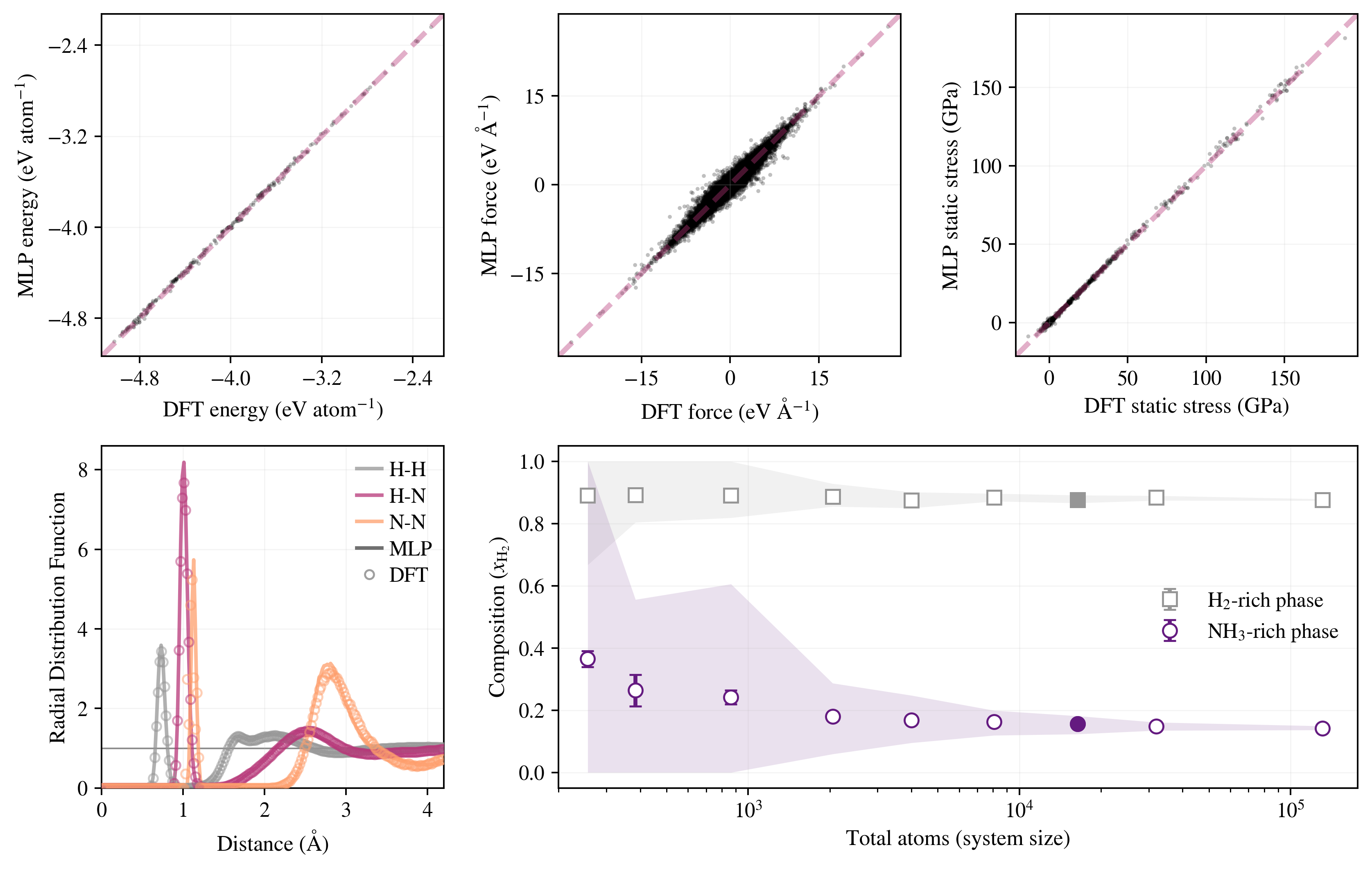}
\caption{Comparison of the machine-learning potential (MLP) predictions with density functional theory (DFT) calculations, and the dependence of our results on system size. The top panels compare the energies, forces and static stresses predicted by the MLP with those from DFT for 61,562 atoms, corresponding to 258 energies, 184,686 force components and 1,548 stress components at 500-4000 K; the dashed lines indicate perfect agreement. The root-mean-square errors are 19 meV atom$^{-1}$ in energy, 0.39 eV \AA$^{-1}$ in force and 0.96 GPa in stress. The bottom left panel compares the H-H, H-N and N-N radial distribution functions from the MLP (solid lines) and from DFT (open circles) at 1000 K and 30 GPa. The bottom right panel shows the compositions of the hydrogen-rich phase (squares) and the ammonia-rich phase (circles) at 1000 K and 10 GPa as a function of the number of atoms in the simulation; the shaded bands show the frame-to-frame fluctuations, and the filled symbols mark the chosen 16,384-atom system.}
\label{fig:parity}
\end{figure*}

\begin{figure*}
\centering
\includegraphics[width=\textwidth,trim=0 0 0 0,clip]{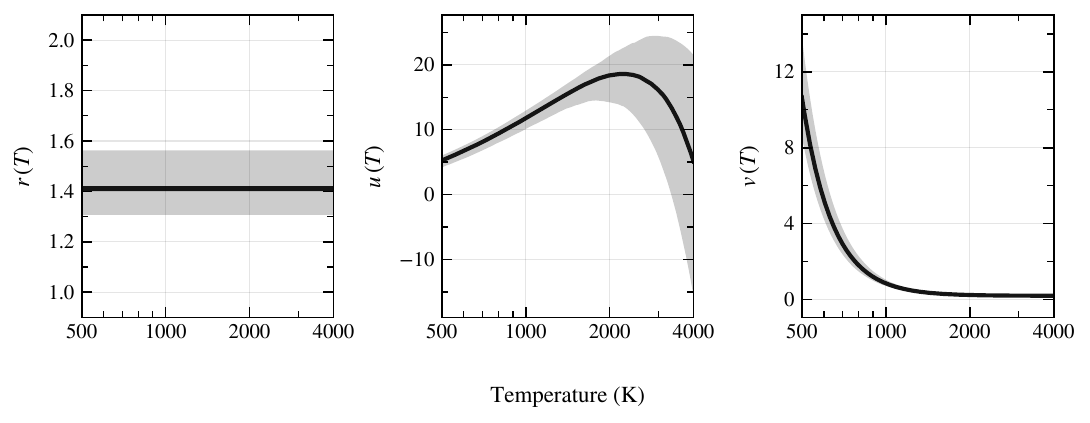}
\caption{Temperature dependence of the mixing parameters of the thermodynamic model (Equation \ref{eq:gs-family}): the dimensionless asymmetry parameter $r$ (left), the zero-pressure excess Gibbs energy scale $u$ in kJ mol$^{-1}$ (middle), and the excess volume scale $v$ in kJ mol$^{-1}$ GPa$^{-1}$ (right), on a logarithmic temperature axis. Solid curves show pointwise posterior medians, and shaded bands 2$\sigma$ intervals, conditional on the model. The selected model has a temperature-independent $r$, while $u$ is quadratic in $t=T/(1000\ {\rm K})$ and $v$ contains a term proportional to $t^{-4}$.}
\label{fig:thermodynamic-functions}
\end{figure*}

\begin{figure}
\centering
\includegraphics[width=0.4\textwidth,trim=0 0 0 0,clip]{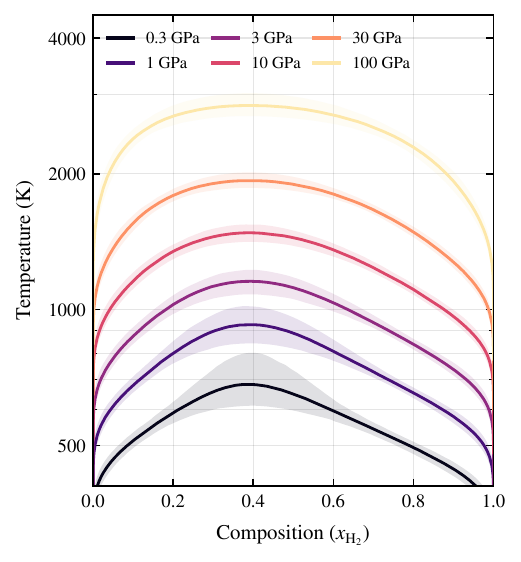}
\caption{Coexistence (binodal) curves predicted by the thermodynamic model at fixed pressures of 0.3-100 GPa, shown as temperature against composition on a logarithmic temperature axis. Below each curve a mixture of that composition separates into two phases. Solid curves show posterior medians, and shaded bands 2$\sigma$ intervals. Portions below 500 K and above 2000 K extend beyond the temperatures with observed two-phase states.}
\label{fig:binodals}
\end{figure}

\end{document}